\documentclass[12pt]{article}
\usepackage{color}
\usepackage{amssymb,amsmath,comment}
\usepackage{graphicx}
\usepackage{braket}
\usepackage{epsfig}
\usepackage{here}
\graphicspath{{./Figures/}}
\newcommand{\ba}{\begin{align}}
\newcommand{\ea}{\end{align}}

\newcommand{\ov}{\overline}
\def\nn{\nonumber}

\def\bea{\begin{eqnarray}}
\def\eea{\end{eqnarray}}
 \makeatletter
\def\alt{\mathrel{\mathpalette\gl@align<}}
\def\agt{\mathrel{\mathpalette\gl@align>}}
\def\gl@align#1#2{\lower.6ex\vbox{\baselineskip\z@skip\lineskip\z@
\ialign{$\m@th#1\hfil##\hfil$\crcr#2\crcr\sim\crcr}}} \makeatother
\renewcommand{\thefootnote}{\fnsymbol{footnote}}
\begin{document}
\begin{flushright}
\end{flushright}
\vspace*{1.0cm}

\begin{center}
\baselineskip 20pt 
{\Large\bf 
Enhanced $\Gamma(p\to K^0\mu^+)/\Gamma(p\to K^+\bar{\nu}_\mu)$ as a Signature of
\\
Minimal Renormalizable SUSY $SO(10)$ GUT
}
\vspace{1cm}

{\large 
Naoyuki Haba$^a$,
\ Yukihiro Mimura$^{a,b}$
\ and \ Toshifumi Yamada$^a$
} \vspace{.5cm}

{\baselineskip 20pt \it
$^a$ Institute of Science and Engineering, Shimane University, Matsue 690-8504, Japan\\
$^b$ Department of Physical Sciences, College of Science and Engineering, \\
Ritsumeikan University, Shiga 525-8577, Japan
}

\vspace{.5cm}

\vspace{1.5cm} {\bf Abstract} \end{center}

The ratio of the partial widths of some dimension-5 proton decay modes can be
  predicted without detailed knowledge of SUSY particle masses,
 and thus allows us to experimentally test various SUSY GUT models without discovering SUSY particles.
In this paper, we study the ratio of the partial widths of the $p\to K^0\mu^+$ and $p\to K^+\bar{\nu}_\mu$ decays
 in the minimal renormalizable SUSY $SO(10)$ GUT, 
 under only a plausible assumption that the 1st and 2nd generation left-handed squarks are mass-degenerate.
In the model, we expect that the Wilson coefficients of dimension-5 operators responsible for these modes
 are on the same order and that the ratio of $p\to K^0\mu^+$ and $p\to K^+\bar{\nu}_\mu$ partial widths is $O(0.1)$.
Hence, we may be able to detect both $p\to K^0\mu^+$ and $p\to K^+\bar{\nu}_\mu$ decays at Hyper-Kamiokande,
 thereby gaining a hint for the minimal renormalizable SUSY $SO(10)$ GUT.
Moreover, since this partial width ratio is quite suppressed in the minimal $SU(5)$ GUT,
 it allows us to distinguish the minimal renormalizable SUSY $SO(10)$ GUT from the minimal $SU(5)$ GUT.
In the main body of the paper, we perform a fitting of the quark and lepton masses and flavor mixings
 with the Yukawa couplings of the minimal renormalizable $SO(10)$ GUT,
 and derive a concrete prediction for the partial width ratio based on the fitting results.
We find that the partial width ratio generally varies in the range 0.05-0.6, confirming the above expectation.

\thispagestyle{empty}

\newpage
\renewcommand{\thefootnote}{\arabic{footnote}}
\setcounter{footnote}{0}
\baselineskip 18pt
\section{Introduction}

The $SO(10)$ grand unified theory (GUT)~\cite{Georgi:1974my,Fritzsch:1974nn} 
 is a well-motivated scenario beyond the Standard Model (SM), since
 it unifies the SM gauge groups into an anomaly-free group, 
 it unifies the SM matter fields and right-handed neutrino of each generation
 into one {\bf 16} representation, and it includes the seesaw mechanism~\cite{seesaw1,seesaw2,seesaw3,seesaw4} for the tiny neutrino mass.
The minimal renormalizable $SO(10)$ GUT~\cite{Babu:1992ia}, 
 where the electroweak-symmetry-breaking-Higgs field stems from ${\bf 10}+{\bf \ov{126}}$
 fields and the SM Yukawa couplings come solely from renormalizable terms 
 $\tilde{Y}_{10}\,{\bf 16}\, {\bf 10}\, {\bf 16}+\tilde{Y}_{126}\,{\bf 16}\, {\bf \ov{126}}\, {\bf 16}$,
 is even more appealing because the mass and flavor mixings of quarks and leptons are derived from a restricted set of parameters.
Specifically, the up-type quark, down-type quark, charged lepton and neutrino Dirac Yukawa matrices are derived as
$Y_u=Y_{10}+r_2Y_{126}$, $Y_d=r_1(Y_{10}+Y_{126})$, $Y_e=r_1(Y_{10}-3Y_{126})$, $Y_D=Y_{10}-3r_2Y_{126}$,
 with $Y_{10}\propto \tilde{Y}_{10}$, $Y_{126}\propto \tilde{Y}_{126}$ and $r_1,r_2$ being numbers.
Also, the Majorana mass for right-handed neutrinos and the type-2 seesaw contribution to the tiny neutrino mass
 are proportional to $Y_{126}$.

The direct experimental signature of the minimal renormalizable $SO(10)$ GUT is, like other GUT models, proton decay.
In supersymmetric (SUSY) GUT, proton decay through dimension-5 operators induced by colored Higgsino exchange~\cite{Weinberg:1981wj,Sakai:1981pk}
 can be within the reach of Hyper-Kamiokande experiment~\cite{Abe:2018uyc}
\footnote{
If ${\bf 45}+{\bf 16}+{\bf \ov{16}}$ fields are responsible for breaking $SO(10)$ gauge group,
 then proton decay through dimension-6 operators induced by GUT gauge boson exchange can also be within the reach of Hyper-Kamiokande~\cite{Haba:2019wwt}.
}
 and is crucial to phenomenology.
Regrettably, SUSY particles have not been discovered at the LHC and hence no concrete prediction is available for the partial widths of dimension-5 proton decays, 
 since they are inversely proportional to the soft SUSY breaking scale squared.
In this situation, the ratio of the partial widths of different decay modes, which is independent of the soft SUSY breaking scale
\footnote{
If the ratio involves a decay mode that receives contributions from both left-handed dimension-5 operators $QQQL$ and right-handed ones $EUUD$,
 we need information about the ratio of Wino mass and $\mu$-term to predict the ratio.
}, allows us to test various SUSY GUT models including the minimal renormalizable SUSY $SO(10)$ GUT.

In this paper, we focus on the ratio of the partial widths of the $p\to K^0 \mu^+$ and the $p\to K^+\bar{\nu}_\mu$ decays
 in the minimal renormalizable SUSY $SO(10)$ GUT.
We make only one natural assumption on the SUSY particle mass spectrum, which is that the 1st and 2nd generation left-handed squarks are mass-degenerate.
In the model with the above assumption, the ratio $\Gamma(p\to K^0\mu^+)/\Gamma(p\to K^+\bar{\nu}_\mu)$
 is predicted to be $O(0.1)$.
Hence, we may be able to discover both $p\to K^0 \mu^+$ and $p\to K^+\bar{\nu}_\mu$ decays at Hyper-Kamiokande~\cite{Abe:2018uyc}, thereby gaining a hint for the model.
Moreover, this ratio is predicted to be suppressed by factor 0.002 in the minimal $SU(5)$ GUT compared to 
 the minimal renormalizable SUSY $SO(10)$ GUT
\footnote{
The origin of the suppression factor 0.002 is explained in Section~\ref{estimate-su5}.
},
 and thus observation of both $p\to K^0 \mu^+$ and $p\to K^+\bar{\nu}_\mu$ decays allows us to distinguish the latter from the former.

In the main body of the paper, we numerically confirm that $\Gamma(p\to K^0\mu^+)/\Gamma(p\to K^+\bar{\nu}_\mu)$ is $O(0.1)$
 in the minimal renormalizable SUSY $SO(10)$ GUT.
To this end, we determine the fundamental Yukawa couplings $Y_{10},Y_{126}$ through a fitting of the quark and lepton Yukawa couplings and neutrino data,
  as has been performed in Refs.~\cite{Matsuda:2000zp}-\cite{Deppisch:2018flu},
 and calculate the partial width ratio based on the fitting results.

Previously, enhancement of partial width ratio $\Gamma(p\to K^0 \mu^+)/\Gamma(p\to K^+\bar{\nu}_\mu)$ in $SO(10)$ GUT models compared to the minimal $SU(5)$ GUT
 is claimed in Refs.~\cite{Babu:1997js,Babu:1998wi},
 but only based on a qualitative argument.
Our paper is the first study where this ratio is predicted concretely and quantitatively in the minimal renormalizable SUSY $SO(10)$ GUT,
 with the fundamental Yukawa couplings $Y_{10},Y_{126}$ determined through a numerical fitting.

The basic reason that $\Gamma(p\to K^0\mu^+)/\Gamma(p\to K^+\bar{\nu}_\mu)$ is $O(0.1)$
 in the minimal renormalizable SUSY $SO(10)$ GUT is understood as follows.
In the model, the ratio of the Wilson coefficients of
 dimension-5 operators responsible for the $p\to K^0\mu^+$ decay and those for the $p\to K^+\bar{\nu}_\mu$ decay,
 is proportional to
 $(Y_{10})_{u_L \,j}/(Y_{10})_{d_L \,j}$ or $(Y_{126})_{u_L \,j}/(Y_{126})_{d_L \,j}$.
Here $(Y_{10})_{u_L \,j}$ denotes (1,$j$)-component of $Y_{10}$ in the flavor basis where,
 when we write the Yukawa coupling as $\psi_i (Y_{10})_{ij}\psi_j$,
 the left-handed up-type quark component of $\psi_i $ has the diagonalized up-type quark Yukawa coupling.
$(Y_{10})_{d_L \,j},(Y_{126})_{u_L \,j},(Y_{126})_{d_L \,j}$ are defined in the same way.
$Y_{10},Y_{126}$ are linear combinations of the down-type and up-type quark Yukawa matrices $Y_d,Y_u$,
 due to the relations $Y_u=Y_{10}+r_2Y_{126},Y_d=r_1(Y_{10}+Y_{126})$.
Moreover, these linear combinations are generic, because situations where $Y_{10}\propto Y_u,Y_{126}\propto Y_d$
 or $Y_{10}\propto Y_d,Y_{126}\propto Y_u$
 would not reproduce the correct charged lepton Yukawa matrix $Y_e$.
Therefore, considering the large hierarchy $y_u/y_t\ll y_d/y_b$, 
 we expect that the components $(Y_{10})_{u_L \,j},(Y_{10})_{d_L \,j},(Y_{126})_{u_L \,j},(Y_{126})_{d_L \,j}$
 are {\it all} on the order of the down quark Yukawa coupling $y_d$ times the mixing angle between the right-handed down quark and 
 a state with flavor index $j$,
 and are {\it not} proportional to the up quark Yukawa coupling $y_u$.
Hence, both $(Y_{10})_{u_L \,j}/(Y_{10})_{d_L \,j}$ and $(Y_{126})_{u_L \,j}/(Y_{126})_{d_L \,j}$ are $O(1)$
 and so is the ratio of the Wilson coefficients of dimension-5 operators for the $p\to K^0\mu^+$ and the $p\to K^+\bar{\nu}_\mu$ decays.
The Wino-dressing diagrams give almost the same contribution for the two modes,
 if the 1st and 2nd generation left-handed squarks are mass-degenerate.
As a result, the partial width ratio $\Gamma(p\to K^0\mu^+)/\Gamma(p\to K^+\bar{\nu}_\mu)$
 is determined by the ratio of baryon chiral Lagrangian parameters,
 which lies in the range $(1-D+F)^2/(1+D+F)^2= 0.085$ to $(1-D+F)^2/(1-D/3+F)^2= 0.30$,
 and thus the partial width ratio is $O(0.1)$.

This paper is organized as follows.
In Section~2, we describe the minimal renormalizable SUSY $SO(10)$ GUT and present formulas for
 the partial widths of the $p\to K^+\bar{\nu}_\mu$ and $p\to K^0 \mu^+$ decays.
In Section~3, we roughly estimate the partial width ratio $\Gamma(p\to K^0\mu^+)/\Gamma(p\to K^+\bar{\nu}_\mu)$
 in the minimal renormalizable SUSY $SO(10)$ GUT without numerically determining the fundamental Yukawa couplings $Y_{10},Y_{126}$,
 and compare it to the partial width ratio in the minimal $SU(5)$ GUT.
In Section~4, we numerically determine $Y_{10},Y_{126}$ through a fitting of the quark and charged lepton Yukawa couplings and neutrino mass matrix,
 and calculate $\Gamma(p\to K^0\mu^+)/\Gamma(p\to K^+\bar{\nu}_\mu)$ based on the fitting results.
Section~5 summarizes the paper.
\\

\section{Minimal Renormalizable SUSY $SO(10)$ GUT}

We consider a SUSY $SO(10)$ GUT model that contains chiral superfields $H$, $\Delta$, $\overline{\Delta}$
 in ${\bf 10}$, ${\bf 126}$, ${\bf \overline{126}}$ representation,
 and three matter fields $\Psi_i$ in {\bf 16} representation ($i=1,2,3$ denotes flavor index)~\cite{Babu:1992ia}.
The model also contains chiral superfields responsible for breaking $SU(5)$ subgroup of $SO(10)$, but we do not specify them in this paper.
The most general renormalizable Yukawa couplings are given by
\bea
W_{\rm Yukawa}\ =\ (\tilde{Y}_{10})_{ij}\,\Psi_i H\Psi_j+(\tilde{Y}_{126})_{ij}\,\Psi_i\overline{\Delta}\Psi_j
\eea
 where $(\tilde{Y}_{10})_{ij}$ and $(\tilde{Y}_{126})_{ij}$ are $3\times3$ complex symmetric matrices.
The Higgs fields of the minimal SUSY Standard Model (MSSM), $H_u,H_d$, are linear combinations of 
 (${\bf 1}$, ${\bf2}$, $\pm\frac{1}{2}$) components of $H$, $\overline{\Delta}$ and other fields.
Accordingly, the MSSM Yukawa coupling for up-type quarks, $Y_u$, that for down-type quarks, $Y_d$, and that for charged leptons, $Y_e$, 
 and the Dirac Yukawa coupling for neutrinos, $Y_D$, are derived from $W_{\rm Yukawa}$ as
\bea
W_{\rm Yukawa}\ \supset\ (Y_u)_{ij}\,Q_i H_u U_i^c+(Y_d)_{ij}\,Q_i H_d D_i^c+(Y_e)_{ij}\,L_i H_d E_i^c+(Y_D)_{ij}\,L_i H_u N_i^c
\label{susyyukawa}
\eea
 where $Y_u,\ Y_d,\ Y_e,\ Y_D$ are given by
\bea
&&Y_u\ =\ Y_{10}+r_2\,Y_{126},
\label{yu}\\
&&Y_d\ =\ r_1\left(Y_{10}+Y_{126}\right),
\label{yd}\\
&&Y_e\ =\ r_1\left(Y_{10}-3Y_{126}\right),
\label{ye}\\
&&Y_D\ =\ Y_{10}-3r_2\,Y_{126}
\label{ydirac}
\eea
 at a $SO(10)$ breaking scale.
Here $Y_{10}\propto\tilde{Y}_{10}$, $Y_{126}\propto\tilde{Y}_{126}$,
 and $r_1,r_2$ are numbers.
By a phase redefinition, we take $r_1$ to be real positive.
In principle, $r_1,r_2$ are determined from the mass matrix for (${\bf 1}$, ${\bf2}$, $\pm\frac{1}{2}$) components~\cite{Fukuyama:2004xs}-\cite{Bajc:2005qe},
 but in this paper we treat them as independent parameters.

Majorana mass for the right-handed neutrinos is proportional to $(Y_{126})_{ij}\,v_R\,N_i^c N_j^c$
 where $v_R$ denotes $\overline{\Delta}$'s VEV.
Integrating out $N_i^c$ yields an effective operator $L_iH_uL_jH_u$, which we call the Type-1 seesaw contribution.
Additionally, if the ({\bf 1}, {\bf3}, 1) component of $\overline{\Delta}$ mixes with that of {\bf 54} representation field,
 after integrating out these components, we get an effective operator
 $L_iH_uL_jH_u$, which we call the Type-2 seesaw contribution.

$H$, $\bar{\Delta}$ and other fields contain pairs of
 ({\bf 3}, {\bf1}, $-\frac{1}{3}$), (${\bf \overline{3}}$, {\bf1}, $\frac{1}{3}$) components,
 which we call `colored Higgs fields' and denote by $H_C^A$, $\ov{H}_C^B$ ($A,B$ are labels), respectively.
Exchange of $H_C^A,\ov{H}_C^B$ gives rise to dimension-5 operators inducing proton decay.
Those couplings of $H_C^A,\overline{H}^B_C$ which contribute to such operators are
\bea
W_{\rm Yukawa}\ \supset\ \sum_A\left[ \ \frac{1}{2}(Y_L^A)_{ij}\,Q_i H_C^A Q_j +  (\overline{Y}^A_L)_{ij}\,Q_i \overline{H}^A_C L_j + (Y_R^A)_{ij}\,E^c_i H_C^A U^c_j +  (\overline{Y}_R^A)_{ij}\,U^c_i \overline{H}_C^A D^c_j  \ \right]
\label{coloredHiggsYukawa}
\eea
where $Y_L^A,\,\ov{Y}_L^A,\,Y_R^A,\,\ov{Y}^A_R$ are linear combinations of $Y_{10},\,Y_{126}$.
After integrating out $H^A_C,\overline{H}^B_C$, we get dimension-5 operators contributing to proton decay,
\bea
-W_5 \ = \ \frac{1}{2}C_{5L}^{ijkl}\,(Q_k Q_{l})(Q_i L_{j}) + C_{5R}^{ijkl}\,E^c_k U^c_{l} U^c_i D^c_{j}
\eea
(in the first term, isospin indices are summed in each bracket) where
\begin{align}
C_{5L}^{ijkl}(\mu\sim M_{H_C})&= \left. \sum_{A,B}({\cal M}_{H_C}^{-1})_{AB}
\left\{(Y_L^A)_{kl}(\ov{Y}^B_L)_{ij}
-\frac{1}{2}(Y_L^A)_{li}(\ov{Y}^B_L)_{kj}-\frac{1}{2}(Y_L^A)_{ik}(\ov{Y}^B_L)_{lj}\right\}\right|_{\mu\sim M_{H_C}},
\label{c5lgeneral}\\
C_{5R}^{ijkl}(\mu\sim M_{H_C})&= \left. \sum_{A,B}({\cal M}_{H_C}^{-1})_{AB}
\left\{(Y_R^A)_{kl}(\ov{Y}_R^B)_{ij}-(Y_R^A)_{ki}(\ov{Y}_R^B)_{lj}\right\}\right|_{\mu\sim M_{H_C}},
\label{c5rgeneral}
\end{align}
 and ${\cal M}_{H_C}$ denotes the mass matrix of $H^A_C,\overline{H}^B_C$ fields and 
 $M_{H_C}$ represents a typical value of the eigenvalues of ${\cal M}_{H_C}$.

We concentrate on the contribution of the $(Q_k Q_{l})(Q_i L_{j})$ operators to the $p\to K^+\bar{\nu}_\mu$ and $p\to K^0\mu^+$ decays,
 and calculate the ratio of their partial widths
\bea
\frac{\Gamma(p\to K^0\mu^+)}{\Gamma(p\to K^+\bar{\nu}_\mu)}
\eea
 in the minimal renormalizable SUSY $SO(10)$ GUT.
It should be noted that
 the $(Q_k Q_{l})(Q_i L_{j})$ and the $E^c_k U^c_{l} U^c_i D^c_{j}$ operators contribute to the $p\to K^+\bar{\nu}_\tau$ decay,
 which is experimentally indistinguishable from the $p\to K^+\bar{\nu}_\mu$ decay.
Hence, our prediction on $\Gamma(p\to K^0\mu^+)/\Gamma(p\to K^+\bar{\nu}_\mu)$
  should be regarded as the maximum of the following measurable quantity:
\bea
  \frac{\Gamma(p\to K^0\mu^+)}{\sum_{i=e,\mu,\tau}\Gamma(p\to K^+\bar{\nu}_i)}
 \eea
The maximum is attained if the $(Q_k Q_{l})(Q_i L_{j})$ operators' contribution and the $E^c_k U^c_{l} U^c_i D^c_{j}$ operators' contribution
  to the $p\to K^+\bar{\nu}_\tau$ decay cancel each other.
This cancellation is always possible by adjusting the ratio of the Wino mass and the $\mu$-term.

As stated in Introduction, for the SUSY particle mass spectrum,
 we assume that the 1st and 2nd generation left-handed squarks are mass-degenerate.
To be quantitative, we assume that the 1st and 2nd generation left-handed squark masses in the up-quark-Yukawa-diagonal basis satisfy
\bea
|m_{\tilde{c}_L}^2 - m_{\tilde{u}_L}^2| \ < \ 10^{-3} \ m_{\tilde{c}_L}^2.
\label{degeneracy}
\eea
This is a natural assumption at the quantum level, since the 1st and 2nd generation quark Yukawa couplings are tiny.
To see this, note that the difference in the renormalization group corrections is given in the leading-log approximation by
\bea
\Delta m_{\tilde{c}_L}^2-\Delta m_{\tilde{u}_L}^2 \ \simeq \ -\frac{3}{16\pi^2}\log\left(\frac{\Lambda^2}{m^2}\right)
\left\{ y_c^2  - y_u^2 + (Y_dY_d^\dagger)_{c_Lc_L} - (Y_dY_d^\dagger)_{u_Lu_L} \right\} m^2
\eea
  where $m^2$ represents the typical scale of soft SUSY breaking masses, and $\Lambda$ denotes the scale at which initial values of the squark masses are given.
We have $|y_c^2  - y_u^2 + (Y_dY_d^\dagger)_{c_Lc_L} - (Y_dY_d^\dagger)_{u_Lu_L}|<10^{-3}$ for $\tan\beta=50$ and at any renormalization scale.
Hence, we get $|\Delta m_{\tilde{c}_L}^2-\Delta m_{\tilde{u}_L}^2|<1.3\times 10^{-3}\ m^2$ even when $\Lambda$ is the Planck scale and $m$ is 1~TeV.
The tiny mass splitting assumed in Eq.~(\ref{degeneracy}) does not affect the results presented in the rest of the paper.

The contribution of the $C_{5L}^{ijkl}(Q_k Q_{l})(Q_i L_{j})$ term to the $p\to K^+\bar{\nu}_\mu$ and the $p\to K^0\mu^+$ decays is given by~\cite{Hisano:1992jj}
\begin{align}
&\Gamma(p\to K^+\bar{\nu}_\mu)
\ = \ {\cal C}
\left\vert \beta_H(\mu_{\rm had})\frac{1}{f_\pi}
\left\{
\left(1+\frac{D}{3}+F\right)C_{LL}^{s\mu \,du}(\mu_{\rm had})+
\frac{2D}{3}C_{LL}^{d\mu \,su}(\mu_{\rm had})\right\}\right\vert^2,
\label{partial1}
\\
&\Gamma(p\to K^0\mu^+)
\ = \ {\cal C}
\left\vert \beta_H(\mu_{\rm had})\frac{1}{f_\pi}
\left(1-D+F\right)\ov{C}_{LL}^{u\mu \,us}(\mu_{\rm had})
\right\vert^2
\label{partial2}
\end{align}
  where ${\cal C} = \frac{m_N}{64\pi}\left(1-\frac{m_{K}^2}{m_N^2}\right)^2$,
  $\beta_H$ denotes a hadronic matrix element, $D,F$ are parameters of the baryon chiral Lagrangian, and
  $C_{LL},\ov{C}_{LL}$ are Wilson coefficients of the effective Lagrangian
  $-{\cal L}_6\supset C_{LL}^{ijkl}(\psi_{u_L^k}\psi_{d_L^l})(\psi_{d_L^i}\psi_{\nu_L^j})+\ov{C}_{LL}^{ijkl}(\psi_{d_L^k} \psi_{u_L^l})(\psi_{u_L^i}\psi_{e_L^j})$ ($\psi$ denotes a SM Weyl fermion and spinor index is summed in each bracket). 
We have neglected the mass splittings among nucleons and hyperons.
The Wilson coefficients $C_{LL},\ov{C}_{LL}$ satisfy
  \footnote{
  When writing $C_{5L}^{u\mu\, us}$, we mean that $Q_i$ is in the flavor basis where the up-type quark Yukawa coupling $Y_u$ is
  diagonal and that the up-type quark component of $Q_i$ is exactly $u$ quark (then the down-type quark component of $Q_i$ is a mixture of $d,s,b$).
  Likewise, $Q_k$ is in the flavor basis where the down-type quark Yukawa coupling $Y_d$ is
  diagonal and its down-type quark component is exactly $s$ quark, 
  and $Q_l$ is in the flavor basis where the up-type quark Yukawa coupling is
  diagonal and its up-type quark component is exactly $u$ quark.
  The same rule applies to other Wilson coefficients.
  }
\begin{align}
C_{LL}^{s\mu\,du}(\mu_{\rm had})= A_{LL}(\mu_{\rm had},\mu_{\rm SUSY})
\frac{M_{\widetilde{W}}}{m_{\tilde{q}}^2}
{\cal F}\,
g_2^2\left(C_{5L}^{s\mu\, ud}-C_{5L}^{u\mu\, sd}\right)\vert_{\mu=\mu_{\rm SUSY}},
\label{cll1}
\\
C_{LL}^{d\mu\,su}(\mu_{\rm had})= A_{LL}(\mu_{\rm had},\mu_{\rm SUSY})
\frac{M_{\widetilde{W}}}{m_{\tilde{q}}^2}
{\cal F}\,
g_2^2\left(C_{5L}^{d\mu\, us}-C_{5L}^{u\mu\, ds}\right)\vert_{\mu=\mu_{\rm SUSY}},
\label{cll2}
\\
\ov{C}_{LL}^{u\mu\,us}(\mu_{\rm had})= A_{LL}(\mu_{\rm had},\mu_{\rm SUSY})
\frac{M_{\widetilde{W}}}{m_{\tilde{q}}^2}
{\cal F}\,
g_2^2\left(-C_{5L}^{u\mu\, us}+C_{5L}^{s\mu\, uu}\right)\vert_{\mu=\mu_{\rm SUSY}},
\label{cll3}
\end{align}
 where ${\cal F}$ is a common loop function factor
 ${\cal F} =\frac{1}{x-y}(\frac{x}{1-x}\log x - \frac{y}{1-y}\log y)/16\pi^2 + \frac{1}{x-1}(\frac{x}{1-x}\log x+1)/16\pi^2$
  with $x=|M_{\widetilde{W}}|^2/m_{\tilde{q}}^2$ and $y=m_{\tilde{\ell}}^2/m_{\tilde{q}}^2$,
  and $m_{\tilde{q}}$ denotes the 1st and 2nd generation left-handed squark masses (which are assumed to be degenerate)
  and $m_{\tilde{\ell}}$ denotes the mass of the left-handed smuon and muon sneutrino.
$A_{LL}(\mu_{\rm had},\mu_{\rm SUSY})$ accounts for renormalization group (RG) corrections in the evolution 
\footnote{
RG corrections involving SM Yukawa couplings are negligible for $C_{LL}^{s\mu\,du},C_{LL}^{d\mu\,su},C_{LL}^{u\mu\,us}$,
 and hence their RG corrections are approximately flavor-universal.
 }
 from soft SUSY breaking scale $\mu_{\rm SUSY}$ to a hadronic scale where the value of $\beta_H$ is reported.
$C_{5L}$ are related to the colored Higgs Yukawa couplings as
\begin{align}
&C_{5L}^{s\mu \,ud}(\mu_{\rm SUSY})-C_{5L}^{u\mu \,sd}(\mu_{\rm SUSY})= 
\nn\\
&\ \ \ \ \ \ \ \ A_L(\mu_{\rm SUSY},\mu_{H_C})
\sum_{A,B}({\cal M}_{H_C}^{-1})_{AB}\,
\frac{3}{2}
\left\{(Y_L^A)_{ud}(\ov{Y}_L^B)_{s\mu}-(Y_L^A)_{ds}(\ov{Y}_L^B)_{u\mu}\right\}|_{\mu=\mu_{H_C}},
\label{c5l1}
\\
&C_{5L}^{d\mu \,us}(\mu_{\rm SUSY})-C_{5L}^{u\mu \,ds}(\mu_{\rm SUSY})=
\nn\\ 
& \ \ \ \ \ \ \ \ A_L(\mu_{\rm SUSY},\mu_{H_C})\sum_{A,B}({\cal M}_{H_C}^{-1})_{AB}\,
\frac{3}{2}
\left\{(Y_L^A)_{us}(\ov{Y}_L^B)_{d\mu}-(Y_L^A)_{ds}(\ov{Y}_L^B)_{u\mu}\right\}|_{\mu=\mu_{H_C}},
\label{c5l2}
\\
&C_{5L}^{u\mu \,us}(\mu_{\rm SUSY})-C_{5L}^{s\mu \,uu}(\mu_{\rm SUSY})= 
\nn\\
& \ \ \ \ \ \ \ \ A_L(\mu_{\rm SUSY},\mu_{H_C})\sum_{A,B}({\cal M}_{H_C}^{-1})_{AB}\,
\frac{3}{2}
\left\{(Y_L^A)_{us}(\ov{Y}_L^B)_{u\mu}-(Y_L^A)_{uu}(\ov{Y}_L^B)_{s\mu}\right\}|_{\mu=\mu_{H_C}},
\label{c5l3}
\end{align}
where $A_L(\mu_{\rm SUSY},\mu_{H_C})$ accounts for RG corrections in the evolution from
colored Higgs mass scale $\mu_{H_C}\sim M_{H_C}$ to soft SUSY breaking scale $\mu_{\rm SUSY}$ 
\footnote{
Again, RG corrections involving MSSM Yukawa couplings are negligible for 
 $C_{5L}^{d\mu \,us}$, $C_{5L}^{u\mu \,ds}$, $C_{5L}^{u\mu \,su}$, $C_{5L}^{s\mu \,uu}$
 and hence their RG corrections are approximately flavor-universal.
 }.

We relate the flavor-dependent part of Eqs.~(\ref{c5l1})-(\ref{c5l3}) to $Y_{10},Y_{126}$.
Since $Y_L^A,\ov{Y}_L^A$ are proportional to either $Y_{10}$ or $Y_{126}$, we can write without loss of generality
\begin{align}
&\sum_{A,B}({\cal M}_{H_C}^{-1})_{AB}\left\{(Y_L^A)_{ud}(\ov{Y}^B_L)_{s\mu}-(Y_L^A)_{ds}(\ov{Y}^B_L)_{u\mu}\right\}
\nn\\
&=\frac{1}{M_{H_C}}
\left[
a\left\{(Y_{10})_{u_Ld_L}(Y_{10})_{s_L\mu_L}-(Y_{10})_{d_Ls_L}(Y_{10})_{u_L\mu_L}\right\}
+b\left\{(Y_{10})_{u_Ld_L}(Y_{126})_{s_L\mu_L}-(Y_{10})_{d_Ls_L}(Y_{126})_{u_L\mu_L}\right\}\right.
\nn\\
&\left.+c\left\{(Y_{126})_{u_Ld_L}(Y_{10})_{s_L\mu_L}-(Y_{126})_{d_Ls_L}(Y_{10})_{u_L\mu_L}\right\}
+d\left\{(Y_{126})_{u_Ld_L}(Y_{126})_{s_L\mu_L}-(Y_{126})_{d_Ls_L}(Y_{126})_{u_L\mu_L}\right\}\right]
\label{yyl1}
\\
&\sum_{A,B}({\cal M}_{H_C}^{-1})_{AB}\left\{(Y_L^A)_{us}(\ov{Y}^B_L)_{d\mu}-(Y_L^A)_{ds}(\ov{Y}^B_L)_{u\mu}\right\}
\nn\\
&=\frac{1}{M_{H_C}}
\left[
a\left\{(Y_{10})_{u_Ls_L}(Y_{10})_{d_L\mu_L}-(Y_{10})_{d_Ls_L}(Y_{10})_{u_L\mu_L}\right\}
+b\left\{(Y_{10})_{u_Ls_L}(Y_{126})_{d_L\mu_L}-(Y_{10})_{d_Ls_L}(Y_{126})_{u_L\mu_L}\right\}\right.
\nn\\
&\left.+c\left\{(Y_{126})_{u_Ls_L}(Y_{10})_{d_L\mu_L}-(Y_{126})_{d_Ls_L}(Y_{10})_{u_L\mu_L}\right\}
+d\left\{(Y_{126})_{u_Ls_L}(Y_{126})_{d_L\mu_L}-(Y_{126})_{d_Ls_L}(Y_{126})_{u_L\mu_L}\right\}\right]
\label{yyl2}
\\
&\sum_{A,B}({\cal M}_{H_C}^{-1})_{AB}\left\{(Y_L^A)_{us}(\ov{Y}^B_L)_{u\mu}-(Y_L^A)_{uu}(\ov{Y}^B_L)_{s\mu}\right\}
\nn\\
&=\frac{1}{M_{H_C}}
\left[
a\left\{(Y_{10})_{u_Ls_L}(Y_{10})_{u_L\mu_L}-(Y_{10})_{u_L u_L}(Y_{10})_{s_L\mu_L}\right\}
+b\left\{(Y_{10})_{u_Ls_L}(Y_{126})_{u_L\mu_L}-(Y_{10})_{u_L u_L}(Y_{126})_{s_L\mu_L}\right\}\right.
\nn\\
&\left.+c\left\{(Y_{126})_{u_L s_L}(Y_{10})_{u_L\mu_L}-(Y_{126})_{u_L u_L}(Y_{10})_{s_L\mu_L}\right\}
+d\left\{(Y_{126})_{u_L s_L}(Y_{126})_{u_L\mu_L}-(Y_{126})_{u_L u_L}(Y_{126})_{s_L\mu_L}\right\}\right]
\label{yyl3}
\end{align}
where $M_{H_C}$ is a typical value of the eigenvalues of ${\cal M}_{H_C}$, and
 $a,b,c,d$ are numbers common for Eqs.~(\ref{yyl1})-(\ref{yyl3}).
Here $(Y_{10})_{u_Ls_L}$ denotes $(1,2)$-component of $Y_{10}$ of the term
 $(Y_{10})_{ij}\,\Psi_i H \Psi_j$
 in the flavor basis where the left-handed up-type quark component of $\Psi_i$ has the diagonalized up-type quark Yukawa coupling,
 and the left-handed down-type quark component of $\Psi_j$ has the diagonalized down-type quark Yukawa coupling.
$(Y_{10})_{d_L\mu_L}$, $(Y_{126})_{u_L s_L}$ and others are defined analogously.

In principle, numbers $a,b,c,d$ are determined from the colored Higgs mass matrix~\cite{Fukuyama:2004xs}-\cite{Bajc:2005qe}.
However, as we do not specify fields responsible for breaking $SU(5)$ subgroup of $SO(10)$,
 we treat $a,b,c,d$ as independent $O(1)$ parameters.

We observe that each term in Eq.~(\ref{yyl3}) is given by $(Y_{10})_{u_L\,j}/(Y_{10})_{d_L\,j}$ or $(Y_{126})_{u_L\,j}/(Y_{126})_{d_L\,j}$
 times some term in Eqs.~(\ref{yyl1}),(\ref{yyl2}), as advertised in Introduction.
For example, the term $(Y_{10})_{u_Ls_L}(Y_{10})_{u_L\mu_L}$ in Eq.~(\ref{yyl3}) 
 equals $(Y_{10})_{u_Ls_L}/(Y_{10})_{d_Ls_L}$ times the term $(Y_{10})_{d_Ls_L}(Y_{10})_{u_L\mu_L}$ in Eq.~(\ref{yyl1}),
 and also equals $(Y_{10})_{u_L\mu_L}/(Y_{10})_{d_L\mu_L}$ times the term $(Y_{10})_{u_Ls_L}(Y_{10})_{d_L\mu_L}$ in Eq.~(\ref{yyl2}).
\\

\section{Estimates on $\Gamma(p\to K^0\mu^+)/\Gamma(p\to K^+\bar{\nu}_\mu)$}

We estimate $\Gamma(p\to K^0\mu^+)/\Gamma(p\to K^+\bar{\nu}_\mu)$ in the minimal $SU(5)$ GUT and in the minimal renormalizable SUSY $SO(10)$ GUT 
 without numerically determining $Y_{10},Y_{126}$.
In the minimal $SU(5)$ GUT, we assume, as usual, that the splitting between the down-type quark Yukawa coupling $Y_d$ and the charged lepton Yukawa coupling $Y_e$
 is realized by non-renormalizable terms.
\\

\subsection{Estimate in the Minimal $SU(5)$ GUT}

In the minimal $SU(5)$ GUT, we have only one pair of colored Higgs fields, and $Y_L$ and $\ov{Y}_L$ are proportional to
 the Yukawa couplings for ${\bf 5}$ and ${\bf \ov{5}}$ Higgs fields, respectively.
Hence, Eqs.~(\ref{yyl1})-(\ref{yyl3}) are altered to
\begin{align}
&\sum_{A,B}({\cal M}_{H_C}^{-1})_{AB}\left\{(Y_L^A)_{ud}(\ov{Y}^B_L)_{s\mu}-(Y_L^A)_{ds}(\ov{Y}^B_L)_{u\mu}\right\}
=\frac{1}{M_{H_C}}
\left\{(Y_{5})_{u_Ld_L}(Y_{\ov{5}})_{s_L\mu_L}-(Y_{5})_{d_Ls_L}(Y_{\ov{5}})_{u_L\mu_L}\right\}
\label{yyl1-alt}
\\
&\sum_{A,B}({\cal M}_{H_C}^{-1})_{AB}\left\{(Y_L^A)_{ud}(\ov{Y}^B_L)_{s\mu}-(Y_L^A)_{ds}(\ov{Y}^B_L)_{u\mu}\right\}
=\frac{1}{M_{H_C}}
\left\{(Y_{5})_{u_Ls_L}(Y_{\ov{5}})_{d_L\mu_L}-(Y_{5})_{d_Ls_L}(Y_{\ov{5}})_{u_L\mu_L}\right\}
\label{yyl2-alt}
\\
&\sum_{A,B}({\cal M}_{H_C}^{-1})_{AB}\left\{(Y_L^A)_{us}(\ov{Y}^B_L)_{u\mu}-(Y_L^A)_{uu}(\ov{Y}^B_L)_{s\mu}\right\}
=\frac{1}{M_{H_C}}
\left\{(Y_{5})_{u_Ls_L}(Y_{\ov{5}})_{u_L\mu_L}-(Y_{5})_{u_L u_L}(Y_{\ov{5}})_{s_L\mu_L}\right\}
\label{yyl3-alt}
\end{align}
 where $Y_{5}$ and $Y_{\ov{5}}$ denote the Yukawa couplings for ${\bf 5}$ and ${\bf \ov{5}}$ Higgs fields, respectively,
 and $M_{H_C}$ denotes the mass for the colored Higgs fields.

The key fact is that since $Y_{5}$ is identical to the up-type quark Yukawa coupling matrix, 
 the components of $Y_{5}$ with flavor index $u_L$ 
 are given by the up quark Yukawa coupling times a mixing angle.
Hence, they are estimated to be
\bea
(Y_{5})_{u_Lu_L}, \ (Y_{5})_{u_Ld_L} &\sim& y_u(\mu=\mu_{H_C})
\\
(Y_{5})_{u_Ls_L} &\sim& y_u(\mu=\mu_{H_C})\cdot \lambda
\eea
 where $\mu_{H_C}\sim M_{H_C}$, and 
 $\lambda$ denotes the Cabibbo angle $\lambda \simeq |V_{us}|\simeq |V_{cd}| \simeq0.22$.
On the other hand, $(Y_{5})_{d_L s_L}$ is estimated to be the second generation Yukawa coupling times a mixing angle as
\bea
(Y_{5})_{d_L s_L} \ \sim \ y_c(\mu_{H_C})\cdot\lambda
\eea
Although the unification of down-type quark Yukawa coupling and charged lepton Yukawa coupling is unsuccessful at the renormalizable level 
 (but the unification can always be achieved with non-renormalizable terms),
 we can estimate components of $Y_{\ov{5}}$ as
\bea
(Y_{\ov{5}})_{s_L\mu_L} &\sim& y_s(\mu_{H_C}) \ \ {\rm or } \ \ y_\mu(\mu_{H_C}),
\\
(Y_{\ov{5}})_{u_L\mu_L} \ \sim \ (Y_{\ov{5}})_{d_L\mu_L} &\sim& y_s(\mu_{H_C})\cdot\lambda \ \ {\rm or } \ \ y_\mu(\mu_{H_C})\cdot\lambda.
\label{y5bar}
\eea
From formulas Eqs.~(\ref{partial1})-(\ref{c5l3}) and estimates Eqs.~(\ref{yyl1-alt})-(\ref{y5bar}), 
 we estimate the partial widths as
\footnote{
We neglect the small difference between hyperon masses and the nucleon mass.
}
\begin{align}
&\Gamma(p\to K^+\bar{\nu}_\mu)
\ = \ {\cal C}\,\left| (1+\frac{D}{3}+F)c_1\,\lambda^2\,y_c\,y_\mu
+\frac{2D}{3}c_2\,\lambda^2\,y_c\,y_\mu\right|^2
\label{partial1estimate}
\\
&\Gamma(p\to K^0\mu^+)
\ = \ {\cal C}\,\left| (1-D+F)c_3\,y_u\,y_\mu\right|^2
\label{partial2estimate}
\end{align}
 or
\begin{align}
 &\Gamma(p\to K^+\bar{\nu}_\mu)
\ = \ {\cal C}\,\left| (1+\frac{D}{3}+F)c_1\,\lambda^2\,y_c\,y_s
+\frac{2D}{3}c_2\,\lambda^2\,y_c\,y_s\right|^2
\label{partial1estimate2}
\\
&\Gamma(p\to K^0\mu^+)
\ = \ {\cal C}\,\left| (1-D+F)c_3\,y_u\,y_s\right|^2
\label{partial2estimate2}
\end{align}
 where ${\cal C}$ is a common constant, $c_1,c_2,c_3$ are $O(1)$ numbers,
 and $y_u,y_c,y_\mu,y_s$ are the up, charm, muon and strange quark Yukawa couplings
 at scale $\mu=\mu_{H_C}$. We have discarded subleading terms.
The partial width ratio is then estimated as
\begin{align}
\left(\frac{1-D+F}{1+D+F}\right)^2\left(\frac{y_u}{\lambda^2y_c}\right)^2 \ \lesssim \
\frac{\Gamma(p\to K^0\mu^+)}{\Gamma(p\to K^+\bar{\nu}_\mu)} \ \lesssim \
\left(\frac{1-D+F}{1-D/3+F}\right)^2\left(\frac{y_u}{\lambda^2y_c}\right)^2
\end{align}
 where the variation is due to unknown relative phase between $c_1$ and $c_2$.
Numerically, the above estimate becomes
\begin{align}
\left(\frac{1-D+F}{1+D+F}\right)^2\cdot0.002 \ \lesssim \
\frac{\Gamma(p\to K^0\mu^+)}{\Gamma(p\to K^+\bar{\nu}_\mu)} \ \lesssim \
\left(\frac{1-D+F}{1-D/3+F}\right)^2\cdot0.002.
\label{ratiosu5}
\end{align}
We find that $p\to K^0\mu^+$ partial width is quite suppressed compared to $p\to K^+\bar{\nu}_\mu$ partial width
 because of the factor 0.002 coming from the ratio of $y_u$ and $\lambda^2y_c$,
 namely, the large hierarchy between the up and charm quark Yukawa couplings suppresses the partial width ratio.
Also, baryon chiral Lagrangian parameters give $(1-D+F)^2/(1+D+F)^2=0.085$ and $(1-D+F)^2/(1-D/3+F)^2=0.3$,
 and they provide further suppression.
\label{estimate-su5}
\\

\subsection{Estimate in the Minimal Renormalizable SUSY $SO(10)$ GUT}

In the minimal renormalizable SUSY $SO(10)$ GUT, we can rewrite the right-hand side of Eqs.~(\ref{yyl1})-(\ref{yyl3}) 
 using the relation $Y_u=Y_{10}+r_2\,Y_{126}$, as
\begin{align}
&\sum_{A,B}({\cal M}_{H_C}^{-1})_{AB}\left\{(Y_L^A)_{ud}(\ov{Y}^B_L)_{s\mu}-(Y_L^A)_{ds}(\ov{Y}^B_L)_{u\mu}\right\}
\nn\\
&=\frac{1}{M_{H_C}}
\left[
a\left\{(Y_{u})_{u_Ld_L}(Y_{u})_{s_L\mu_L}-(Y_{u})_{d_Ls_L}(Y_{u})_{u_L\mu_L}\right\}
+b'\left\{(Y_{u})_{u_Ld_L}(Y_{126})_{s_L\mu_L}-(Y_{u})_{d_Ls_L}(Y_{126})_{u_L\mu_L}\right\}\right.
\nn\\
&\left.+c'\left\{(Y_{126})_{u_Ld_L}(Y_{u})_{s_L\mu_L}-(Y_{126})_{d_Ls_L}(Y_{u})_{u_L\mu_L}\right\}
+d'\left\{(Y_{126})_{u_Ld_L}(Y_{126})_{s_L\mu_L}-(Y_{126})_{d_Ls_L}(Y_{126})_{u_L\mu_L}\right\}\right]
\label{yyl1-alt2}
\\
&\sum_{A,B}({\cal M}_{H_C}^{-1})_{AB}\left\{(Y_L^A)_{us}(\ov{Y}^B_L)_{d\mu}-(Y_L^A)_{ds}(\ov{Y}^B_L)_{u\mu}\right\}
\nn\\
&=\frac{1}{M_{H_C}}
\left[
a\left\{(Y_{u})_{u_Ls_L}(Y_{u})_{d_L\mu_L}-(Y_{u})_{d_Ls_L}(Y_{u})_{u_L\mu_L}\right\}
+b'\left\{(Y_{u})_{u_Ls_L}(Y_{126})_{d_L\mu_L}-(Y_{u})_{d_Ls_L}(Y_{126})_{u_L\mu_L}\right\}\right.
\nn\\
&\left.+c'\left\{(Y_{126})_{u_Ls_L}(Y_{u})_{d_L\mu_L}-(Y_{126})_{d_Ls_L}(Y_{u})_{u_L\mu_L}\right\}
+d'\left\{(Y_{126})_{u_Ls_L}(Y_{126})_{d_L\mu_L}-(Y_{126})_{d_Ls_L}(Y_{126})_{u_L\mu_L}\right\}\right]
\label{yyl2-alt2}
\\
&\sum_{A,B}({\cal M}_{H_C}^{-1})_{AB}\left\{(Y_L^A)_{us}(\ov{Y}^B_L)_{u\mu}-(Y_L^A)_{uu}(\ov{Y}^B_L)_{s\mu}\right\}
\nn\\
&=\frac{1}{M_{H_C}}
\left[
a\left\{(Y_{u})_{u_Ls_L}(Y_{u})_{u_L\mu_L}-(Y_{u})_{u_L u_L}(Y_{u})_{s_L\mu_L}\right\}
+b'\left\{(Y_{u})_{u_Ls_L}(Y_{126})_{u_L\mu_L}-(Y_{u})_{u_L u_L}(Y_{126})_{s_L\mu_L}\right\}\right.
\nn\\
&\left.+c'\left\{(Y_{126})_{u_L s_L}(Y_{u})_{u_L\mu_L}-(Y_{126})_{u_L u_L}(Y_{u})_{s_L\mu_L}\right\}
+d'\left\{(Y_{126})_{u_L s_L}(Y_{126})_{u_L\mu_L}-(Y_{126})_{u_L u_L}(Y_{126})_{s_L\mu_L}\right\}\right]
\label{yyl3-alt2}
\end{align}
 where
\bea
b'=b-r_2\,a, \ \ \ c'=c-r_2\,a, \ \ \ d'=d-r_2(b+c)+r_2^2\,a.
\eea
We still have $b',c',d'=O(1)$, since we have $|r_2|=O(1)$ to fit the charged lepton Yukawa coupling.
The right-hand sides of Eqs.~(\ref{yyl1-alt2})-(\ref{yyl3-alt2}) contain terms analogous to Eqs.~(\ref{yyl1-alt})-(\ref{yyl3-alt})
 (note that $Y_{u}$ in Eqs.~(\ref{yyl1-alt2})-(\ref{yyl3-alt2}) corresponds to $Y_5$ in Eqs.~(\ref{yyl1-alt})-(\ref{yyl3-alt})),
 plus non-analogous terms in the form $(Y_{126})_{ij}(Y_{126})_{kl}$.
Each component is estimated as follows.
$(Y_u)_{s_L\mu_L}$ is estimated to be the charm quark Yukawa coupling
 and $(Y_u)_{d_Ls_L}$ is estimated to be the charm quark Yukawa coupling times the Cabibbo angle,
\bea
&&(Y_u)_{s_L\mu_L} \ \sim \ y_c(\mu_{H_C})
\label{yusLmuLestimate}\\
&&(Y_u)_{d_Ls_L} \ \sim \ y_c(\mu_{H_C}) \cdot \lambda
\eea
The components of $Y_u$ with flavor index $u_L$ are always given by the up Yukawa coupling $y_u$ times a mixing angle, and hence we get
\bea
&&(Y_{u})_{u_Lu_L}, \ (Y_{u})_{u_Ld_L} \ \sim \ y_u(\mu_{H_C})
\\
&&(Y_{u})_{u_Ls_L}, \ (Y_{u})_{u_L\mu_L} \ \sim \  y_u(\mu_{H_C})\cdot\lambda
\eea
In contrast, the components of $Y_{126}$ do not follow the rule and are estimated as
\bea
(Y_{126})_{u_Lu_L}, \ (Y_{126})_{u_Ld_L} &\sim& \frac{1}{r_1}y_d(\mu_{H_C})
\label{y126uLdLestimate}\\
(Y_{126})_{u_Ls_L}, \ (Y_{126})_{d_Ls_L}, \ (Y_{126})_{u_L\mu_L}
 &\sim& \frac{1}{r_1}y_s(\mu_{H_C}) \cdot \lambda
 \label{y126uLsLestimate}\\
(Y_{126})_{s_L\mu_L} &\sim& \frac{1}{r_1}y_s(\mu_{H_C})
 \label{y126sLmuLestimate}
\eea
We have estimated $(Y_{126})_{s_L\mu_L}$ to be $y_s(\mu_{H_C})/r_1$, because we empirically have
 $y_\mu/y_s|_{\mu= 10^{16}~{\rm GeV}}\simeq4$ and this factor 4 is mostly explained by the factor 3 in Eq.~(\ref{ye}).
We have estimated $(Y_{126})_{u_Lu_L}$ to be $y_d(\mu_{H_C})/r_1$, not $y_u(\mu_{H_C})$, based on the following argument:
Recall that components of $Y_{10}$ and $Y_{126}$ reproduce the up and down Yukawa couplings as
\bea
(Y_{10})_{u_Ru_L}+r_2(Y_{126})_{u_Ru_L} &=& y_u(\mu_{H_C})
\label{yu11}\\
r_1\left((Y_{10})_{d_Rd_L}+(Y_{126})_{d_Rd_L}\right)&=& y_d(\mu_{H_C})
\label{yd11}
\eea
Since the unification of the top and bottom Yukawa couplings requires $\tan\beta/r_1\simeq m_t/m_b\simeq50$, we get
\bea
\frac{(Y_{10})_{u_Ru_L}+r_2(Y_{126})_{u_Ru_L}}{(Y_{10})_{d_Rd_L}+(Y_{126})_{d_Rd_L}}
&=&r_1\,\frac{y_u}{y_d}\ = \ \frac{r_1}{\tan\beta}\frac{m_u}{m_d} \ \simeq \ \frac{m_b}{m_t}\frac{m_u}{m_d} \ \simeq \ 0.01.
\label{mumd}
\eea
$(Y_{10})_{u_Ru_L}/(Y_{10})_{d_Rd_L}$ and $(Y_{126})_{u_Ru_L}/(Y_{126})_{d_Rd_L}$ are estimated to be $1-\lambda^2\simeq1$.
Then, the only way to realize Eq.~(\ref{mumd}) is to take
\bea
(Y_{10})_{d_Rd_L}\simeq-r_2(Y_{126})_{d_Rd_L}\simeq\frac{1}{r_1}\frac{r_2}{r_2-1}y_d(\mu_{H_C})
\label{y10drdl}
\eea
 and impose a fine-tuning between $(Y_{10})_{u_Ru_L}$ and $r_2(Y_{126})_{u_Ru_L}$
 to realize the small value 0.01 in Eq.~(\ref{mumd}).
Here we cannot assume $r_2\simeq0$ because we need $|r_2|=O(1)$ to reproduce the charged lepton Yukawa coupling,
 as will be confirmed numerically in Fig.~\ref{r2plot}.
From Eq.~(\ref{y10drdl}), we find
\bea
(Y_{10})_{u_Ru_L}\simeq-r_2(Y_{126})_{u_Ru_L}\simeq\frac{1}{r_1}\frac{r_2}{r_2-1}y_d(\mu_{H_C})
\eea
Using $|r_2|=O(1)$, we estimate $(Y_{10})_{u_Lu_L}$, $(Y_{126})_{u_Lu_L}$ as
\bea
(Y_{10})_{u_Lu_L}, \ (Y_{126})_{u_Lu_L} \ \sim \ \frac{1}{r_1}y_d(\mu_{H_C})
\label{y10uLuLestimate}
\eea

From formulas Eqs.~(\ref{partial1})-(\ref{c5l3}) and estimates Eqs.~(\ref{yusLmuLestimate})-(\ref{y126sLmuLestimate}),
 we estimate the partial widths as
\begin{align}
&\Gamma(p\to K^+\bar{\nu}_\mu)
\ = \ {\cal C}\,\left| (1+\frac{D}{3}+F)(a \beta_1\ y_uy_c+b' \beta_2 \ y_cy_s\lambda^2/r_1+c' \beta_3 \ y_cy_s\lambda^2/r_1+d' \beta_4 \ y_s^2\lambda^2/r_1^2)\right.
\nn\\
& \ \ \ \ \ \ \ \ \ \  \ \ \ \ \ \ \ \ \ \ \left.+\frac{2D}{3}(a \gamma_1\ y_uy_c\lambda^2+b' \gamma_2 \ y_cy_s\lambda^2/r_1+c' \gamma_3 \ y_cy_s\lambda^2/r_1+d' \gamma_4 \ y_s^2\lambda^2/r_1^2)\right|^2,
\label{partial1estimate-so10}
\\
&\Gamma(p\to K^0\mu^+)
\ = \ {\cal C}\,\left| (1-D+F)(a \delta_1 \ y_uy_c+b' \delta_2 \ y_uy_s/r_1+c' \delta_3 \ y_cy_s\lambda^2/r_1+d' \delta_4 \ y_s^2\lambda^2/r_1^2)
\right|^2
\label{partial2estimate-so10}
\end{align}
  where ${\cal C}$ is a common constant, $y_u,y_s,y_c$ are the up, strange and charm quark Yukawa couplings at scale $\mu=\mu_{H_C}$, and
 $\beta_1,\beta_2,\beta_3,\beta_4,\gamma_1,\gamma_2,\gamma_3,\gamma_4,\delta_1,\delta_2,\delta_3,\delta_4$ are $O(1)$ numbers.
We have used empirical relation $m_s\lambda^2\simeq m_d$ and let $y_s\lambda^2$ represent both $y_s\lambda^2$ and $y_d$.

In Eqs.~(\ref{partial1estimate-so10}),(\ref{partial2estimate-so10}), 
 $y_s\lambda^2/r_1^2$ and $y_cy_s\lambda^2/r_1$ are much larger than the other terms containing $y_u$.
Hence, in generic cases where $d'=O(1)$ and/or $c'=O(1)$, the partial width ratio is estimated as
\begin{align}
\left(\frac{1-D+F}{1+D+F}\right)^2 \ &\lesssim \ \frac{\Gamma(p\to K^0\mu^+)}{\Gamma(p\to K^+\bar{\nu}_\mu)}
\ \lesssim \ \left(\frac{1-D+F}{1-D/3+F}\right)^2
\label{ratioso10}\\
&{\rm (in \ minimal \ renormalizable \ }SO(10){\rm \ GUT \ with} \ d'=O(1) \ {\rm and/or} \ c'=O(1))\nn
\end{align}
 where the variation is due to unknown relative phases among $\beta_2,\beta_3,\beta_4,\gamma_2,\gamma_3,\gamma_4$.
We find that the suppression factor of 0.002 in Eq.~(\ref{ratiosu5}) is absent in Eq.~(\ref{ratioso10}).
This means that in the minimal renormalizable SUSY $SO(10)$ GUT with $d'=O(1)$ and/or $c'=O(1)$, 
 $\Gamma(p\to K^0\mu^+)/\Gamma(p\to K^+\bar{\nu}_\mu)$ is highly enhanced compared to the minimal $SU(5)$ GUT.

In the non-generic case where $c'$ and $d'$ are both fine-tuned to 0, the partial width ratio is quite suppressed as
\begin{align}
\left(\frac{1-D+F}{1+D+F}\right)^2 \left(\frac{y_u}{\lambda^2y_c}\right)^2
\ &\lesssim \ \frac{\Gamma(p\to K^0\mu^+)}{\Gamma(p\to K^+\bar{\nu}_\mu)}
\ \lesssim \ \left(\frac{1-D+F}{1-D/3+F}\right)^2 \left(\frac{y_u}{\lambda^2y_c}\right)^2
\label{ratioso10-finetuned}\\
&{\rm (in \ minimal \ renormalizable \ }SO(10){\rm \ GUT \ with} \  c'=d'=0)
\nn
\end{align}
 which is the same as in the minimal $SU(5)$ GUT.
This is reasonable because when $c'=d'=0$,
 the contribution of ({\bf 3}, {\bf1}, $-\frac{1}{3}$) fields to dimension-5 proton decay
 is dictated by the up-type quark Yukawa matrix, just as in the minimal $SU(5)$ GUT.

In the next section, we numerically confirm the estimates Eqs.~(\ref{ratioso10}),(\ref{ratioso10-finetuned})
 through a fitting of the quark and lepton masses and flavor mixings in terms of $Y_{10},Y_{126}$.
\\

\section{Numerical Analysis}

\subsection{Overview}

Our first task is to fit the MSSM Yukawa matrices with $Y_{10},Y_{126},r_1,r_2$ through Eqs.~(\ref{yu})-(\ref{ye}),
 and fit the neutrino mass matrix with $Y_{10},Y_{126},r_2$.
When calculating the Type-1 seesaw contribution to the Weinberg operator $L_iH_uL_jH_u$,
 we have to integrate out each right-handed neutrino $N_i^c$ at its respective mass scale.
This requires information on the eigenvalues of $Y_{126}$, but it is obtained only after the fitting is complete.
Hence, it is technically difficult to integrate out each right-handed neutrino separately.
In this paper, therefore, we make an approximation that the three right-handed neutrinos are integrated out at one scale.
Accordingly, the neutrino mass matrix $M_\nu$ is related to $Y_{126}$ and $Y_D$ in Eq.~(\ref{ydirac}) as
\bea
&&(M_\nu)_{ij} \ \propto \ R_{ik} \left\{ r_L(Y_{126})_{kl}+(Y_D)_{km}(Y_{126}^{-1})_{mn}(Y_D)_{ln} \right\} R_{jl},
\label{rij}
\nn
\eea
 where $r_L$ is a complex number that parametrizes the ratio of the Type-1 and Type-2 seesaw contributions, and
 $R_{ij}$ denotes the flavor-dependent RG correction to the coefficient of the Weinberg operator $L_iH_uL_jH_u$ when it
 evolves from a $SO(10)$ breaking scale to electroweak scale.
Since the flavor-dependent RG correction $R_{ij}$ is at most 3\% (see Table~\ref{values}) 
 while the errors of the neutrino data we employ are much larger (see Table~\ref{fitting}),
 we expect that the approximation of integrating out right-handed neutrinos at one scale does not affect the results.

We repeat the above fitting analysis many times and obtain as many fitting results.
We compute $\Gamma(p\to K^+\bar{\nu}_\mu)$ and $\Gamma(p\to K^0\mu^+)$ from each fitting result of $Y_{10},Y_{126},r_1,r_2,r_L$ using
 Eqs.~(\ref{partial1})-(\ref{c5l3}) and Eqs.~(\ref{yyl1-alt2})-(\ref{yyl3-alt2}),
 with coefficients $a,b',c',d'$ treated as independent $O(1)$ parameters.
The fitting results are plotted with respect to the ratio $\Gamma(p\to K^0\mu^+)/\Gamma(p\to K^+\bar{\nu}_\mu)$.
From the plot, we read out the range of the ratio $\Gamma(p\to K^0\mu^+)/\Gamma(p\to K^+\bar{\nu}_\mu)$ predicted by the minimal renormalizable $SO(10)$ GUT.

We assume a benchmark SUSY particle mass spectrum to evaluate the MSSM Yukawa couplings at a $SO(10)$ breaking scale as well as $R_{ij}$,
 and to compute the individual partial widths $\Gamma(p\to K^+\bar{\nu}_\mu)$ and $\Gamma(p\to K^0\mu^+)$.
However, we emphasize that the purpose of this paper is to predict the ratio $\Gamma(p\to K^0\mu^+)/\Gamma(p\to K^+\bar{\nu}_\mu)$,
 which is not much dependent on the SUSY particle mass spectrum 
 due to the cancellations of the RG corrections and the factors coming from Wino-dressing.
\\

\subsection{Procedures}

First, we numerically calculate the MSSM Yukawa matrices $Y_u,Y_d,Y_e$ at scale $\mu=2\cdot10^{16}$~GeV in $\ov{{\rm DR}}$ scheme,
 and the flavor-dependent RG correction to the coefficient of the Weinberg operator $R_{ij}$.
Specifically, we calculate $R_{ij}$ for the evolution from $\mu=2\cdot10^{16}$~GeV to $\mu=M_Z$.
We assume a high-scale split SUSY particle mass spectrum below for concreteness,
\bea
m_{\widetilde{q}}=m_{\widetilde{\ell}}=m_{H^0}=m_{H^\pm}=m_A=2000~{\rm TeV}, \ \ M_{\widetilde{g}}=M_{\widetilde{W}}=\mu_H=100~{\rm TeV},
\ \ \tan\beta \ = \ 50.
\label{massspectrum}
\eea
For the calculation of the quark Yukawa couplings, we adopt the following input values for quark masses and CKM matrix parameters:
The isospin-averaged quark mass and strange quark mass in $\ov{{\rm MS}}$ scheme are obtained from lattice calculations in
 Refs.~\cite{lattice1,lattice2,lattice3,lattice4,lattice5,lattice6} as
 $\frac{1}{2}(m_u+m_d)(2~{\rm GeV})=3.373(80)~{\rm MeV}$ and $m_s(2~{\rm GeV})=92.0(2.1)~{\rm MeV}$.
The up and down quark mass ratio is obtained from an estimate in Ref.~\cite{latticereview} as $m_u/m_d=0.46(3)$.
The $\ov{{\rm MS}}$ charm and bottom quark masses are obtained from QCD sum rule calculations in Ref.~\cite{cb} as
  $m_c(3~{\rm GeV})=0.986-9(\alpha_s^{(5)}(M_Z)-0.1189)/0.002\pm0.010~{\rm GeV}$
  and $m_b(m_b)=4.163+7(\alpha_s^{(5)}(M_Z)-0.1189)/0.002\pm0.014~{\rm GeV}$.
The top quark pole mass is obtained from $t\bar{t}$+jet events measured by ATLAS~\cite{Aad:2019mkw} as
 $M_t=171.1\pm1.2$~GeV.
The CKM mixing angles and CP phase are calculated from the Wolfenstein parameters in the latest CKM fitter result~\cite{ckmfitter}.
For the QCD and QED gauge couplings, we use $\alpha_s^{(5)}(M_Z)=0.1181$ and $\alpha^{(5)}(M_Z)=1/127.95$.
For the lepton and W, Z, Higgs pole masses, we use the values in Particle Data Group~\cite{Tanabashi:2018oca}.

The results are given in terms of the singular values of $Y_u,Y_d,Y_e$ and the CKM mixing angles and CP phase at $\mu=2\cdot10^{16}$~GeV,
 as well as $R_{ij}$ in the flavor basis where $Y_e$ is diagonal ($R_{ij}$ is also diagonal in this basis),
 tabulated in Table~\ref{values}.
For each singular value of $Y_u,Y_d$, we present 1$\sigma$ error that has propagated from experimental error of
 the corresponding input quark mass.
For the CKM mixing angles and CP phase, we present 1$\sigma$ errors that have propagated from experimental errors of the input Wolfenstein parameters.
\begin{table}[H]
\begin{center}
  \caption{The singular values of MSSM Yukawa couplings $Y_u$, $Y_d$, $Y_e$, and the mixing angles and CP phase of CKM matrix,
  at $\mu=2\cdot10^{16}$~GeV in $\ov{{\rm DR}}$ scheme.
  Also shown is the flavor-dependent RG correction $R_{ij}$ for the Weinberg operator (defined in Eq.~(\ref{rij})) in the evolution from 
   $\mu=2\cdot10^{16}$~GeV to $\mu=M_Z$, in the flavor basis where $Y_e$ is diagonal ($R_{ij}$ is also diagonal in this basis).
  For each singular value of the quark Yukawa matrices, we present 1$\sigma$ error that has propagated from experimental error of 
   the corresponding input quark mass,
  and for the CKM parameters, we present 1$\sigma$ errors that have propagated from experimental errors of the input 
   Wolfenstein parameters.}
  \begin{tabular}{|c|c|} \hline
                                      & Value at $\mu=2\cdot10^{16}$~GeV in $\ov{{\rm DR}}$ scheme \\ \hline
    $y_u$           &2.74(14)$\times10^{-6}$\\
    $y_c$           &0.001407(14)\\
    $y_t$            &0.4620(84)\\ \hline
    $y_d$           &0.0002998(94)\\
    $y_s$           &0.00597(14)\\
    $y_b$           &0.3376(19)\\ \hline
    $y_e$           &0.00012486\\
    $y_\mu$           &0.026364\\
    $y_\tau$           &0.50319\\ \hline
    $\cos\theta_{13}^{\rm ckm}\sin\theta_{12}^{\rm ckm}$            & 0.22475(25)\\
    $\cos\theta_{13}^{\rm ckm}\sin\theta_{23}^{\rm ckm}$           & 0.0421(11)\\
    $\sin\theta_{13}^{\rm ckm}$                                                              & 0.00372(22)\\
    $\delta_{\rm km}$~(rad)           &1.147(33)\\ \hline
    $R_{ee}$ & 1.00\\
    $R_{\mu\mu}$ & 1.00\\\
    $R_{\tau\tau}$ & 0.974\\ \hline
  \end{tabular}
  \label{values}
  \end{center}
\end{table}

To facilitate the fitting analysis, we rearrange Eqs.~(\ref{yu})-(\ref{ye}) as follows.
We fix the flavor basis such that the left-handed up-type quark components in both $\Psi_i$ and $\Psi_j$ have the diagonalized up-type quark Yukawa matrix with real positive components.
$Y_d$, which is still symmetric, is then written as
\footnote{
Note that $Y_d$ in Eq.~(\ref{susyyukawa}) is the complex conjugate of $Y_d$ in SM defined as 
 $-{\cal L}=\bar{q}_L Y_d d_R\, i\sigma_2 H^*$.
}
\bea
Y_d=\begin{pmatrix} 
      1 & 0 & 0 \\
      0 & e^{i \,a_2} & 0 \\
      0 & 0 & e^{i \,a_3}\\
   \end{pmatrix}
V_{\rm CKM}^*   
\begin{pmatrix} 
      y_d  \,e^{2i \,b_1} & 0 & 0 \\
      0 & y_s \,e^{2i \,b_2} & 0 \\
      0 & 0 & y_b \,e^{2i \,b_3}\\
   \end{pmatrix}
   V_{\rm CKM}^\dagger
   \begin{pmatrix} 
      1 & 0 & 0 \\
      0 & e^{i \,a_2} & 0 \\
      0 & 0 & e^{i \,a_3}\\
   \end{pmatrix}
   \label{yd3}
\eea
 where $a_2,a_3,b_1,b_2,b_3$ are unknown phases.
In the same flavor basis, $Y_e$ is written from Eqs.~(\ref{yu})-(\ref{ye}) as
\bea
\frac{1}{r_1}Y_e=\frac{4}{1-r_2}\begin{pmatrix} 
      y_u & 0 & 0 \\
      0 & y_c & 0 \\
      0 & 0 & y_t \\
   \end{pmatrix}
   -
   \frac{3+r_2}{1-r_2}\frac{1}{r_1}Y_d
   \label{ye3}
 \eea
 with $Y_d$ given in Eq.~(\ref{yd3}).
We can also write
\bea
&&Y_{126}\propto\frac{1}{r_1}Y_d-\begin{pmatrix} 
      y_u & 0 & 0 \\
      0 & y_c & 0 \\
      0 & 0 & y_t \\
   \end{pmatrix},
\\
&&Y_D=
\begin{pmatrix} 
      y_u & 0 & 0 \\
      0 & y_c & 0 \\
      0 & 0 & y_t \\
   \end{pmatrix}-\frac{4r_2}{1-r_2}
   \left(
   \frac{1}{r_1}Y_d-\begin{pmatrix} 
      y_u & 0 & 0 \\
      0 & y_c & 0 \\
      0 & 0 & y_t \\
   \end{pmatrix}
   \right).
\eea
Finally, we perform the singular value decomposition of $Y_e$ as
\bea
Y_e = U_{eL}\begin{pmatrix} 
      y_e & 0 & 0 \\
      0 & y_\mu & 0 \\
      0 & 0 & y_\tau \\
   \end{pmatrix}
   U_{eR}^\dagger,
 \eea
and calculate the active neutrino mass matrix (up to overall constant) as
\bea
(M_\nu)_{\ell \ell'} \ \propto \ R_{\ell \ell} \ [\ U_{eL}^T(r_L\,Y_{126}+Y_DY_{126}^{-1}Y_D^T)U_{eL} \ ]_{\ell\ell'} \ 
R_{\ell' \ell'}, \ \ \ \ \ \ \ell,\ell'=e,\mu,\tau,
\label{mnumatching2}
\eea
where $\ell,\ell'$ denote flavor indices for the left-handed charged leptons.
From Eq.~(\ref{mnumatching2}), we derive the three neutrino mixing angles $\theta_{12}^{\rm pmns},\theta_{13}^{\rm pmns},\theta_{23}^{\rm pmns}$ and the ratio of the neutrino masses $m_1:m_2:m_3$.

Now we perform the fitting with $Y_{10},Y_{126},r_1,r_2,r_L$. It proceeds as follows.
We fix $y_u,y_c,y_t$ and CKM matrix by the values in Table~\ref{values},
 while we vary $y_d/r_1,\,y_s/r_1,\,y_b/r_1$, unknown phases $a_2,a_3,b_1,b_2,b_3$ in Eq.~(\ref{yd3}) and complex number $r_2$.
Here we eliminate $r_1$ by requiring that the central value of the electron Yukawa coupling $y_e$ be reproduced.
In this way, we try to 
 reproduce the correct values of $y_d,y_s,y_\mu,y_\tau$, $\theta_{12}^{\rm pmns},\theta_{13}^{\rm pmns},\theta_{23}^{\rm pmns}$
 and neutrino mass difference ratio $\Delta m_{21}^2/\Delta m_{32}^2$.
Specifically, we require $y_d,y_s$ to fit within their respective 3$\sigma$ ranges,
 while we do not constrain $y_b$ because $y_b$ can receive sizable SUSY particle and GUT-scale threshold corrections.
Since the experimental errors of $y_\mu,y_\tau$ are tiny, 
 we only require that their reproduced values fit within $\pm$0.1\% ranges of their central values.
We require $\sin^2\theta_{12}^{\rm pmns},\sin^2\theta_{13}^{\rm pmns},\sin^2\theta_{23}^{\rm pmns}$, 
 $\Delta m_{21}^2/\Delta m_{32}^2
 $ to fit within their respective 3$\sigma$ ranges reported by NuFIT~4.1~\cite{Esteban:2018azc,nufit}.
However, we do not constrain the Dirac CP phase $\delta_{\rm pmns}$, since its measurement is still at a primitive stage.
We only consider the normal mass hierarchy case, because we cannot obtain a good fitting with the inverted mass hierarchy.
We have confirmed that our fitting analysis
  always gives small values for $m_1$ that are not in tension with cosmological observations or searches for neutrinoless double-beta decay,
  and hence no constraint is imposed on $\alpha_2,\alpha_3,m_1$.
The constraints are summarized in Table~\ref{fitting}.
\begin{table}[H]
\begin{center}
  \caption{Allowed ranges of quantities in the analysis.}
  \begin{tabular}{|c|c|} \hline
    Quanitity & Allowed range \\ \hline
    $y_u$           &2.74$\times10^{-6}$ \ \ (fixed)\\
    $y_c$           &0.001407 \ \ (fixed)\\
    $y_t$            &0.4620 \ \ (fixed)\\ \hline
    $y_d$           &0.0002998$\pm0.0000094\cdot3$\\
    $y_s$           &0.00597$\pm0.00014\cdot3$\\
    $y_b$           &unconstrained\\ \hline
    $y_e$           &0.00012486 (used to fix $r_1$)\\
    $y_\mu$           &0.026364$\pm 0.1$\%\\
    $y_\tau$           &0.50319$\pm 0.1$\% \\ \hline
    $\cos\theta_{13}^{\rm ckm}\sin\theta_{12}^{\rm ckm}$            & 0.22475 (fixed)\\
    $\cos\theta_{13}^{\rm ckm}\sin\theta_{23}^{\rm ckm}$           & 0.0421 (fixed)\\
    $\sin\theta_{13}^{\rm ckm}$                                                              & 0.00372 (fixed)\\
    $\delta_{\rm km}$~(rad)           &1.147 (fixed)\\ \hline
$\sin^2\theta_{12}^{\rm pmns}$ & $[0.275,~0.350]$\\
    $\sin^2\theta_{13}^{\rm pmns}$ & $[0.02044,~0.02435]$\\
    $\sin^2\theta_{23}^{\rm pmns}$ & $[0.433,~0.609]$\\
    $\Delta m_{21}^2/\Delta m_{32}^2$ & $[0.0267,~0.0339]$ \\
    $\delta_{\rm pmns}, \ \alpha_2,\ \alpha_3,\ m_1$ & unconstrained \\ \hline
    $a_2, \ a_3, \ b_1, \ b_2,\ b_3$ & unconstrained \\
    $r_1$ & eliminated in favor of $y_e$ \\
    $r_2$ & unconstrained \\ \hline
  \end{tabular}
  \label{fitting}
  \end{center}
\end{table}

We collect sets of values of $Y_{10},Y_{126},r_1,r_2,r_L$ that satisfy the constraints of Table~\ref{fitting}.
From these values, we reconstruct the MSSM Yukawa couplings $Y_u,Y_d,Y_e$, perform flavor basis changes, and
 calculate the following components:
\bea
(Y_{u})_{u_Ld_L}, \ \ \ (Y_{u})_{s_L\mu_L}, \ \ \ (Y_{u})_{d_Ls_L}, \ \ \ (Y_{u})_{u_L\mu_L},
\nn\\
(Y_{126})_{u_Ld_L}, \ \ \ (Y_{126})_{s_L\mu_L}, \ \ \ (Y_{126})_{d_Ls_L}, \ \ \ (Y_{126})_{u_L\mu_L}.
\nn
\eea
From the values above, we calculate $\Gamma(p\to K^+\bar{\nu}_\mu)$ and $\Gamma(p\to K^0\mu^+)$
 through Eqs.~(\ref{partial1})-(\ref{c5l3}) and Eqs.~(\ref{yyl1-alt2})-(\ref{yyl3-alt2}),
 by considering various $O(1)$ values for coefficients $a,b',c',d'$ in Eqs.~(\ref{yyl1-alt2})-(\ref{yyl3-alt2}).
Here we take $M_{H_C}=2\cdot10^{16}$~GeV and assume the SUSY particle mass spectrum of Eq.~(\ref{massspectrum}).
We employ the following data and formulas.
For the hadronic matrix element $\beta_H$, we adopt the value in Ref.~\cite{Aoki:2017puj}, which reads $\beta_H=0.0144$~GeV$^3$
 at $\mu=2$~GeV in $\ov{{\rm MS}}$ scheme.
The baryon chiral Lagrangian parameters are given by $D=0.804$, $F=0.463$,
 and we include the mass splittings among nucleon and hyperon masses found in Particle Data Group~\cite{Tanabashi:2018oca}.
When computing RG corrections to the dimension-5 operators and the dimension-6 operators after Wino-dressing,
 we choose $\mu_{\rm SUSY}=2000$~TeV and $\mu_{H_C}=2\cdot10^{16}$~GeV,
 and use one-loop formulas in Ref.~\cite{Goto:1998qg}.
\\

\subsection{Results}

We have obtained 158 sets of values of $Y_{10},Y_{126},r_1,r_2,r_L$ that satisfy the constraints of Table~\ref{fitting}.

\begin{figure}[H]
\begin{center}
\includegraphics[width=110mm]{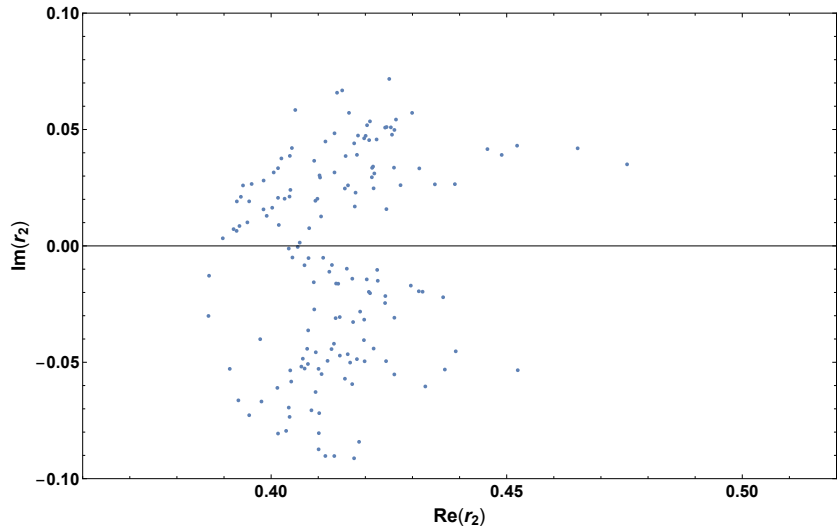}
\caption{
Distribution of $r_2$ (defined in Eq.~(\ref{yu})) in the fitting results satisfying the constraints of Table~\ref{fitting}.
}
\label{r2plot}
\end{center}
\end{figure}

Before presenting the main results, we show in Fig.~\ref{r2plot} the distribution of $r_2$ in the fitting results,
 to confirm the relation $|r_2|=O(1)$ used in Section~3.

Now we plot the sets of values of $Y_{10},Y_{126},r_1,r_2,r_L$ satisfying Table~\ref{fitting},
 on the plane of $p\to K^+\bar{\nu}_\mu$ partial lifetime versus the ratio of the partial widths of $p\to K^0\mu^+$ and $p\to K^+\bar{\nu}_\mu$.
From the plots, we read out the range of the partial width ratio predicted by the model.

\begin{figure}[H]
\begin{center}
\includegraphics[width=80mm]{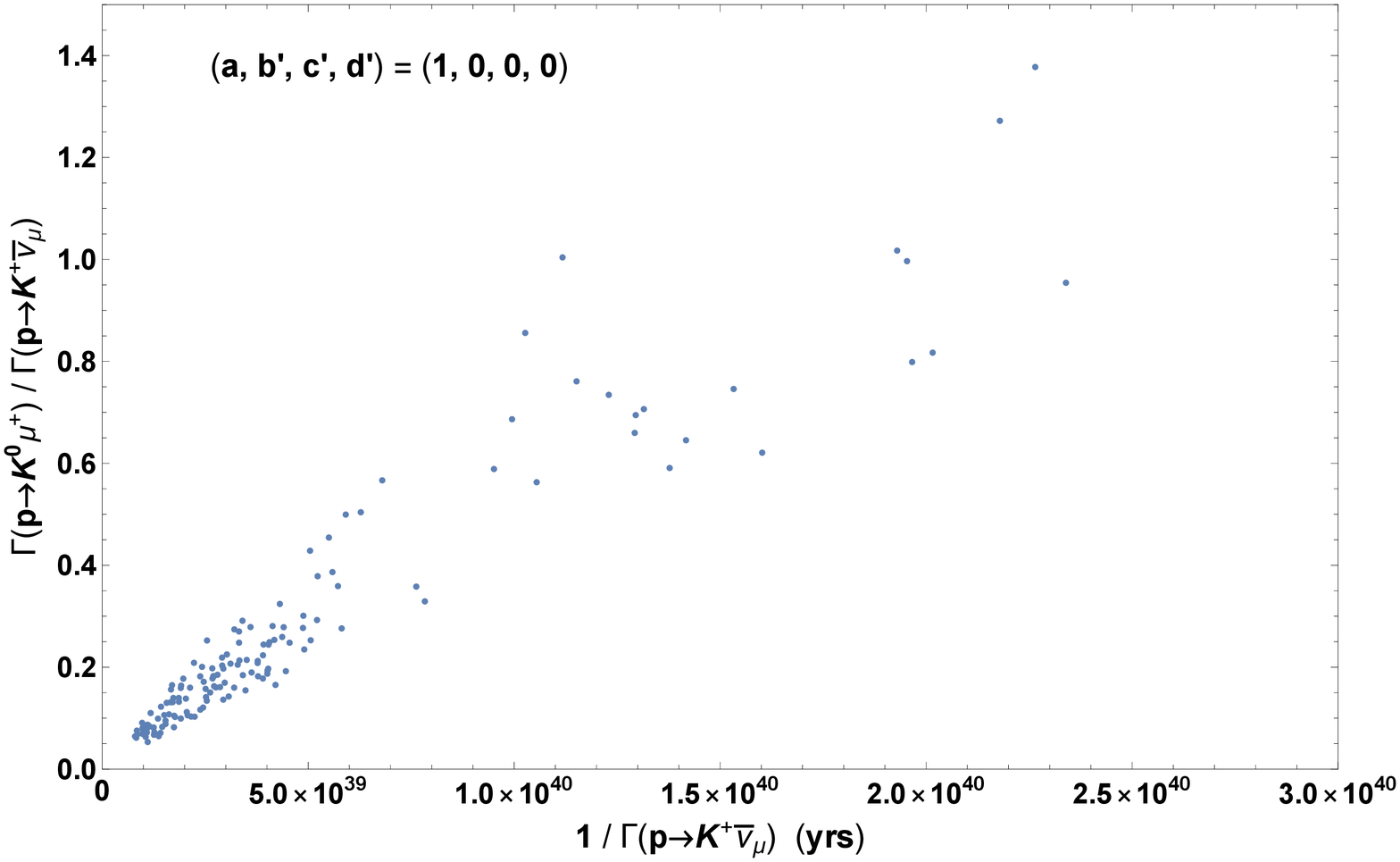}
\includegraphics[width=80mm]{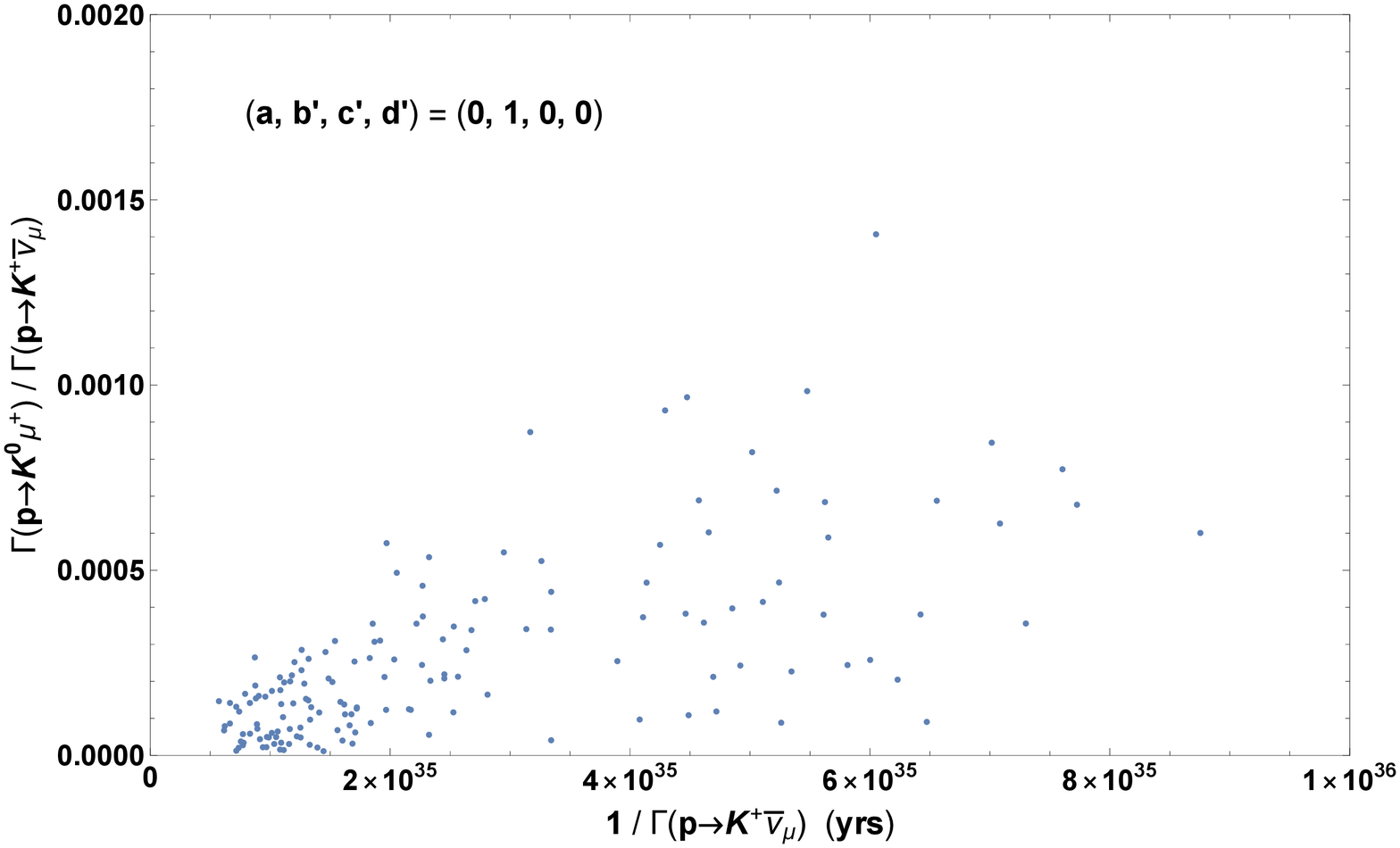}
\\
\includegraphics[width=80mm]{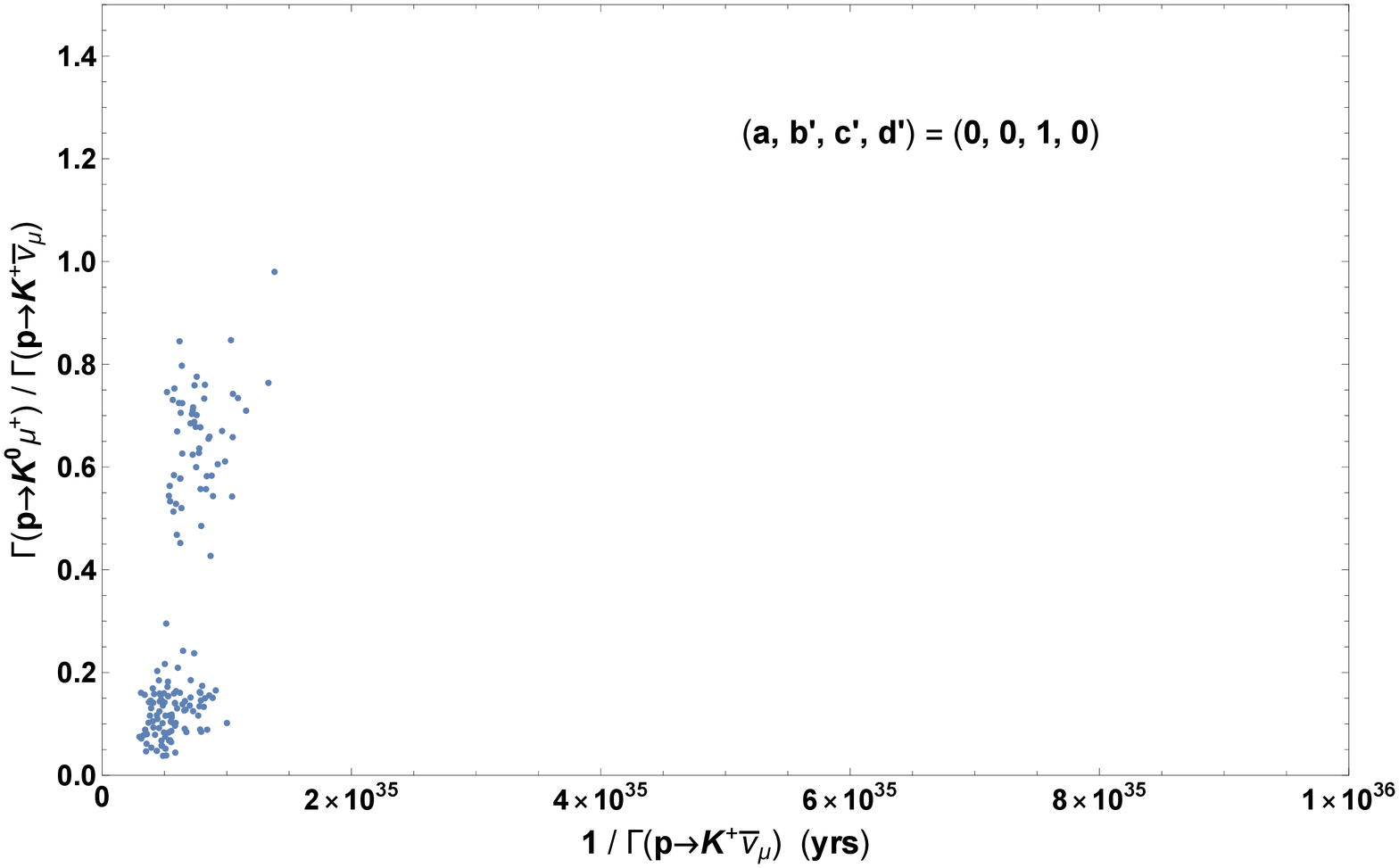}
\includegraphics[width=80mm]{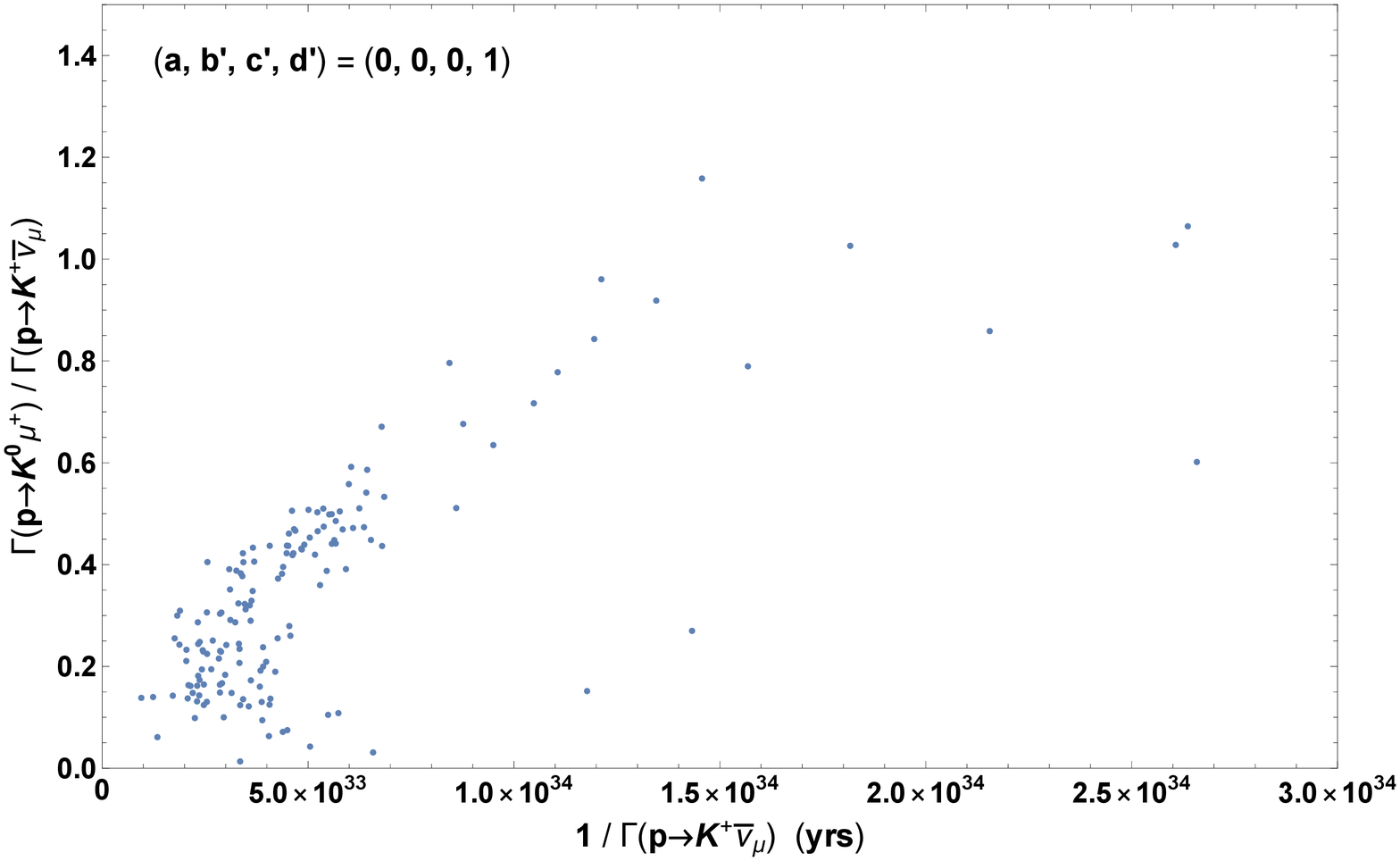}
\caption{
$p\to K^+\bar{\nu}_\mu$ partial lifetime versus the ratio of the partial widths of $p\to K^0\mu^+$ and $p\to K^+\bar{\nu}_\mu$.
Each dot corresponds to a set of values of $Y_{10},Y_{126},r_1,r_2,r_L$ that satisfy the constraints of Table~\ref{fitting}.
We take $(a,b',c',d')=(1,0,0,0), \, (0,1,0,0), \, (0,0,1,0), \, (0,0,0,1)$ in Eq.~(\ref{yyl1-alt2})-(\ref{yyl3-alt2}).
Note that the vertical scale of the panel of $(a,b',c',d')=(0,1,0,0)$ is different because the partial width ratio is quite suppressed in this case.
Also, the horizontal scale is different for the four panels, due to the large hierarchy of $p\to K^+\bar{\nu}_\mu$ partial lifetime in the four cases.
Although some points 
 are apparently excluded by the current 90\% CL experimental bound $1/\Gamma(p\to K^+\nu) > 5.9\times10^{33}$~years~\cite{Abe:2014mwa},
 these points are revived if $(a,b',c',d')$ are reduced due to the mixing of 
 ({\bf 3}, {\bf1}, $-\frac{1}{3}$), (${\bf \overline{3}}$, {\bf1}, $\frac{1}{3}$) components of fields other than $H,\ov{\Delta}$,
 or if SUSY particles are slightly heavier than the spectrum of Eq.~(\ref{massspectrum}).
}
\label{plot-indiv}
\end{center}
\end{figure}

We first study the contribution of individual terms in Eqs.~(\ref{yyl1-alt2})-(\ref{yyl3-alt2})
 by taking $(a,b',c',d')=(1,0,0,0), \, (0,1,0,0), \, (0,0,1,0), \, (0,0,0,1)$.
The plots are in Fig.~\ref{plot-indiv}.
We caution that although some points 
 are apparently excluded by the current 90\% CL experimental bound $1/\Gamma(p\to K^+\nu) > 5.9\times10^{33}$~years~\cite{Abe:2014mwa},
 these points are revived if $(a,b',c',d')$ are reduced due to the mixing of 
 ({\bf 3}, {\bf1}, $-\frac{1}{3}$), (${\bf \overline{3}}$, {\bf1}, $\frac{1}{3}$) components of fields other than $H,\ov{\Delta}$,
 or if SUSY particles are slightly heavier than the spectrum of Eq.~(\ref{massspectrum}) by factor $O(1)$.
We find that the predictions for $\Gamma(p\to K^+\bar{\nu}_\mu)$ in the cases with
 $(a,b',c',d')=(1,0,0,0)$, $(0,1,0,0)$, $(0,0,1,0)$, $(0,0,0,1)$
 exhibit the following hierarchy:
\bea
({\rm case \ with \ }(1,0,0,0))\, \ll \,({\rm case \ with \ }(0,1,0,0))\, \lesssim \,({\rm case \ with \ }(0,0,1,0))\, \ll \,({\rm case \ with \ }(0,0,0,1))\nn
\eea
On the other hand, the predictions for the partial width ratio $\Gamma(p\to K^0\mu^+)/\Gamma(p\to K^+\bar{\nu}_\mu)$
 follow the following pattern:
\bea
({\rm case \ with \ }(0,1,0,0)) \, \ll \, ({\rm case \ with \ }(1,0,0,0)) \, \sim \, 
({\rm case \ with \ }(0,0,1,0)) \, \sim \, ({\rm case \ with \ }(0,0,0,1))\nn
\eea
From the above hierarchy patterns,
  we infer $\Gamma(p\to K^0\mu^+)/\Gamma(p\to K^+\bar{\nu}_\mu)$
 for general values of $(a,b',c',d')$ as follows.
\begin{itemize}

\item
When $d'=O(1)$, the partial width $\Gamma(p\to K^+\bar{\nu}_\mu)$ is dominated by the contribution from
 the term with coefficient $d'$.
Since the partial width ratio $\Gamma(p\to K^0\mu^+)/\Gamma(p\to K^+\bar{\nu}_\mu)$ with $(a,b',c',d')=(0,0,0,1)$
 is comparable to or larger than in the other cases,
 we expect that $\Gamma(p\to K^0\mu^+)$ is also dominated by the contribution from
 the term with $d'$.
Therefore, we conclude that {\it when $d'=O(1)$, irrespectively of the values of $a,b',c'$,
 the prediction on the partial width ratio is given by the lower-right panel of Fig.~\ref{plot-indiv},
 where the partial width ratio mostly varies in the range 0.05-0.6.}
This result is consistent with our estimate Eq.~(\ref{ratioso10}).

\item
When $d'=0$, the partial width $\Gamma(p\to K^+\bar{\nu}_\mu)$ receives comparable contributions from
 the terms with $c'$ and $b'$.
On the other hand, since the partial width ratio $\Gamma(p\to K^0\mu^+)/\Gamma(p\to K^+\bar{\nu}_\mu)$ with $(a,b',c',d')=(0,1,0,0)$
 is much smaller than that with $(a,b',c',d')=(0,0,1,0)$,
 $\Gamma(p\to K^0\mu^+)$ receives contribution solely from the term with $c'$.
Hence, when $c'=O(1)$ and $b'=O(1)$,
  the partial width ratio $\Gamma(p\to K^0\mu^+)/\Gamma(p\to K^+\bar{\nu}_\mu)$ is suppressed
  if the contributions of the terms with $c'$ and $b'$ to 
  $\Gamma(p\to K^+\bar{\nu}_\mu)$ interfere constructively,
  and the partial width ratio is enhanced if they interfere destructively.
To examine these possibilities, we present plots for cases with $(a,b',c',d')=(0,1,1,0)$, $(0,i,1,0)$, $(0,-1,1,0)$, $(0,-i,1,0)$ in Fig.~\ref{plot-cb}.
\begin{figure}[H]
\begin{center}
\includegraphics[width=80mm]{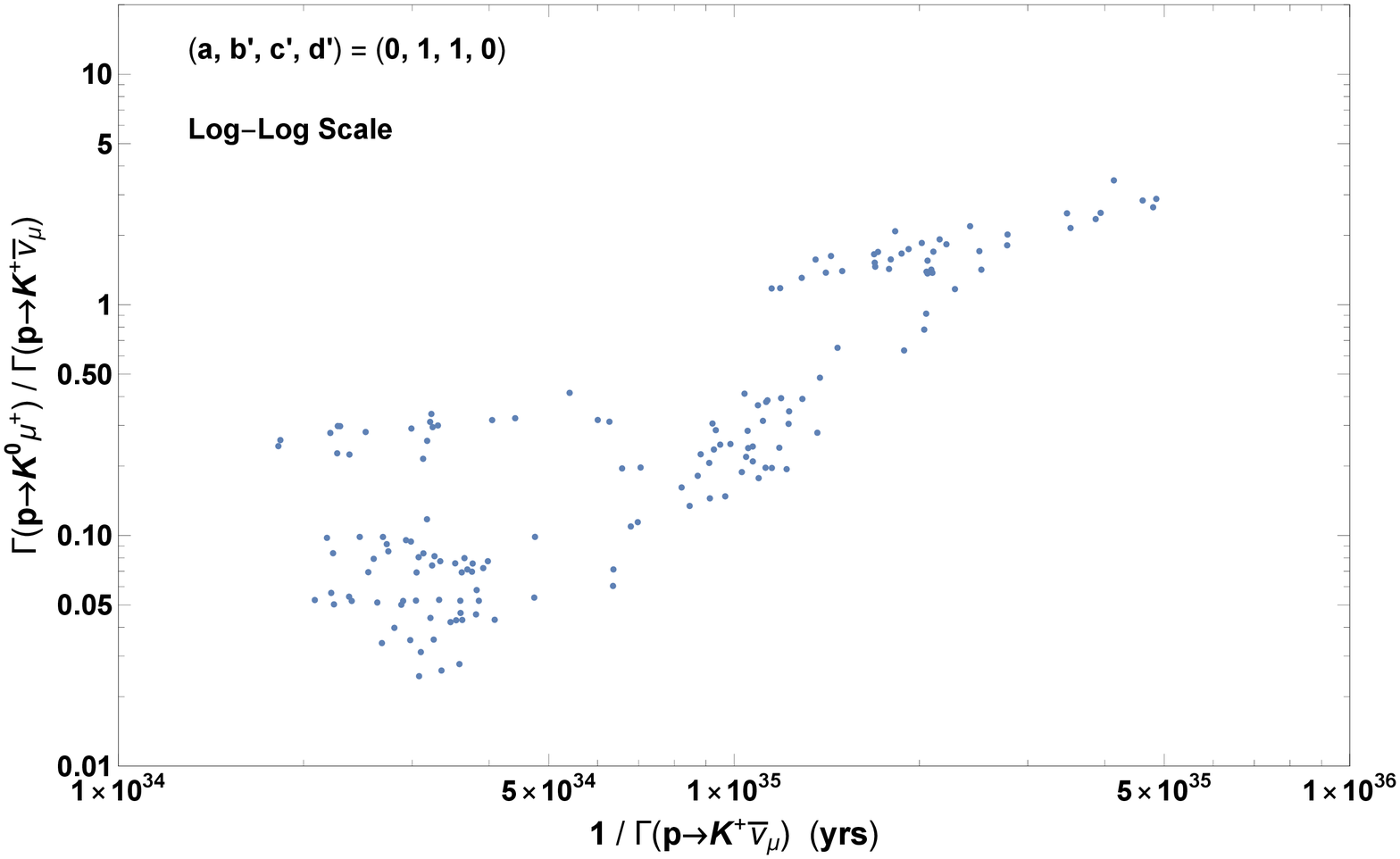}
\includegraphics[width=80mm]{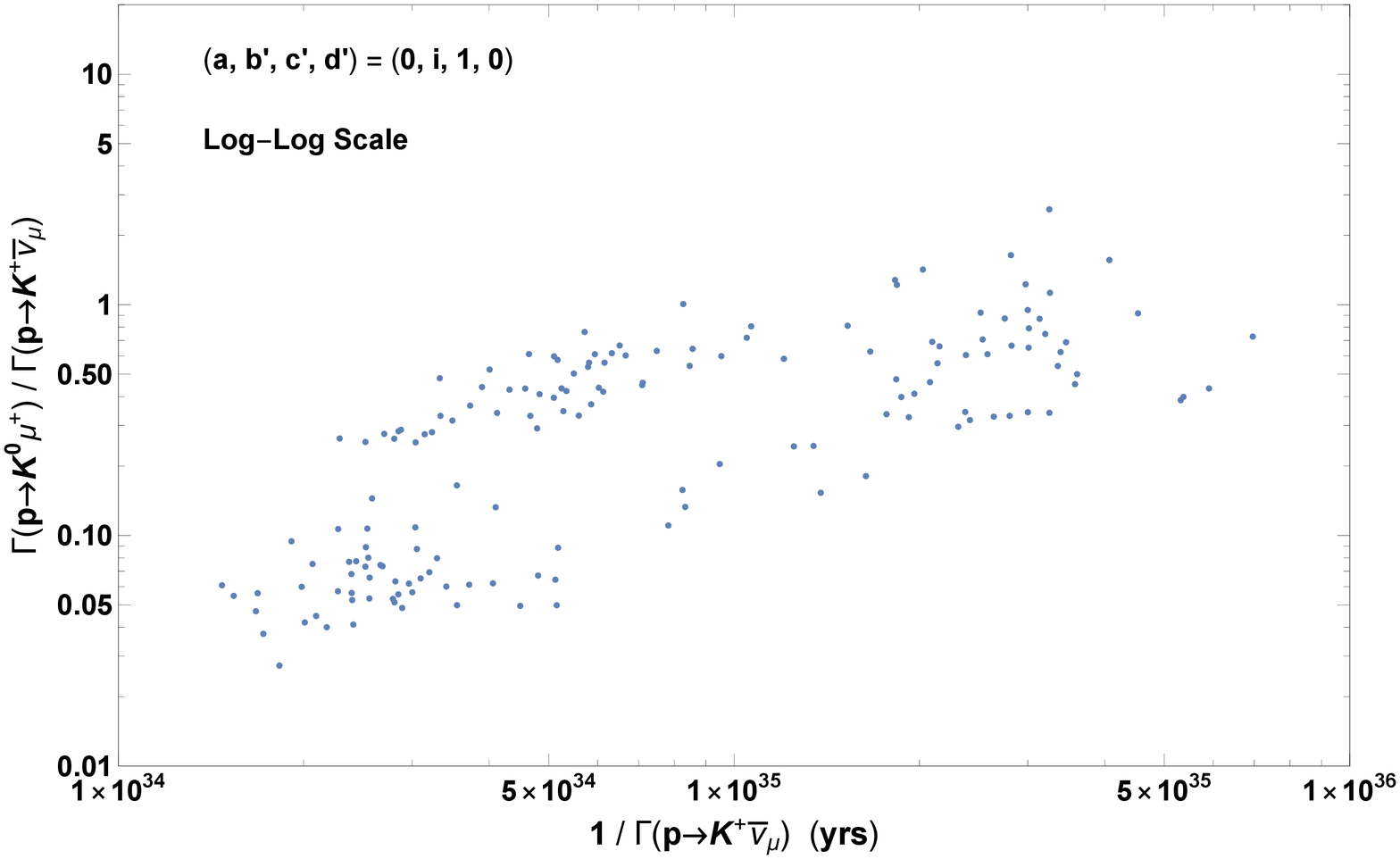}
\\
\includegraphics[width=80mm]{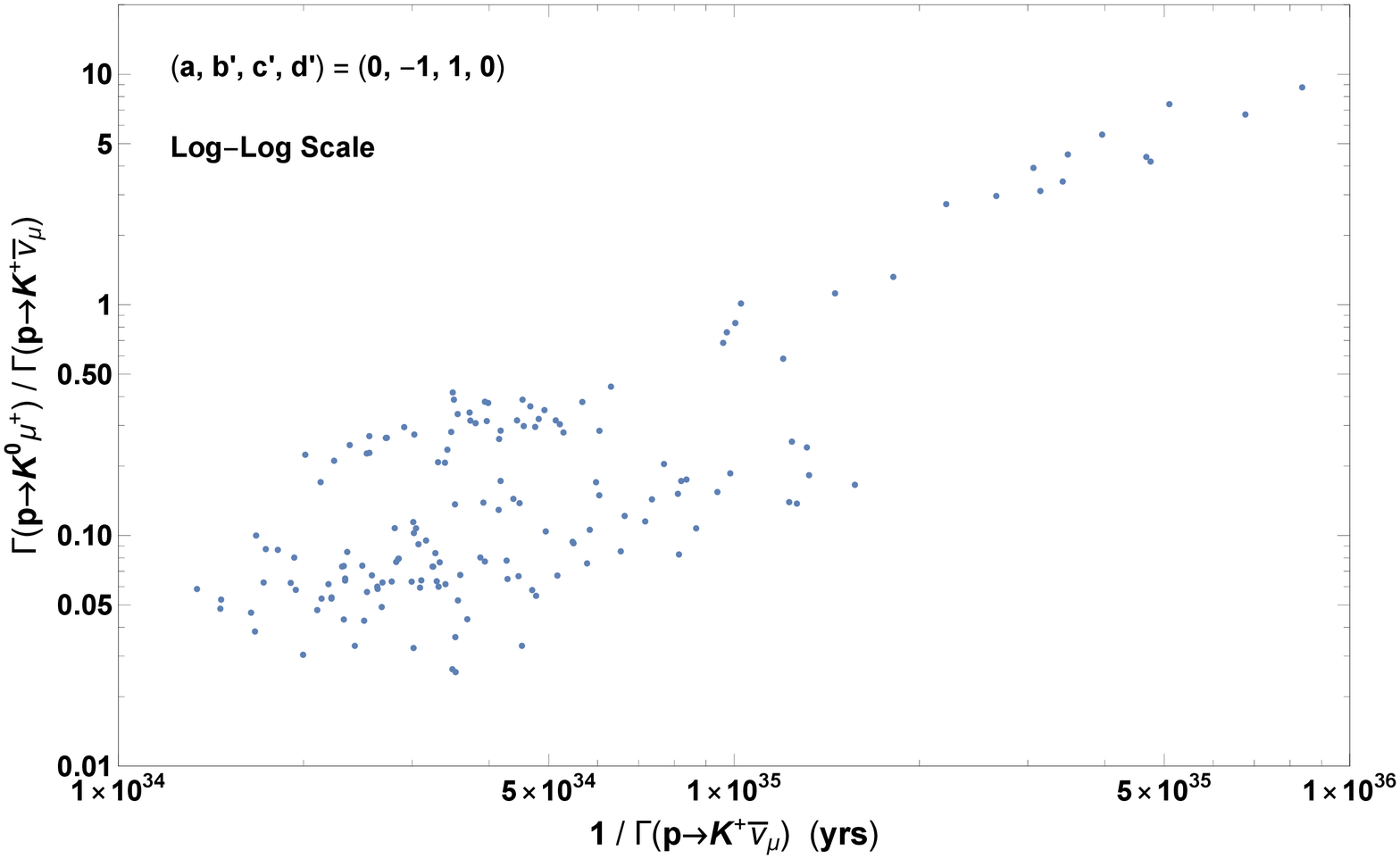}
\includegraphics[width=80mm]{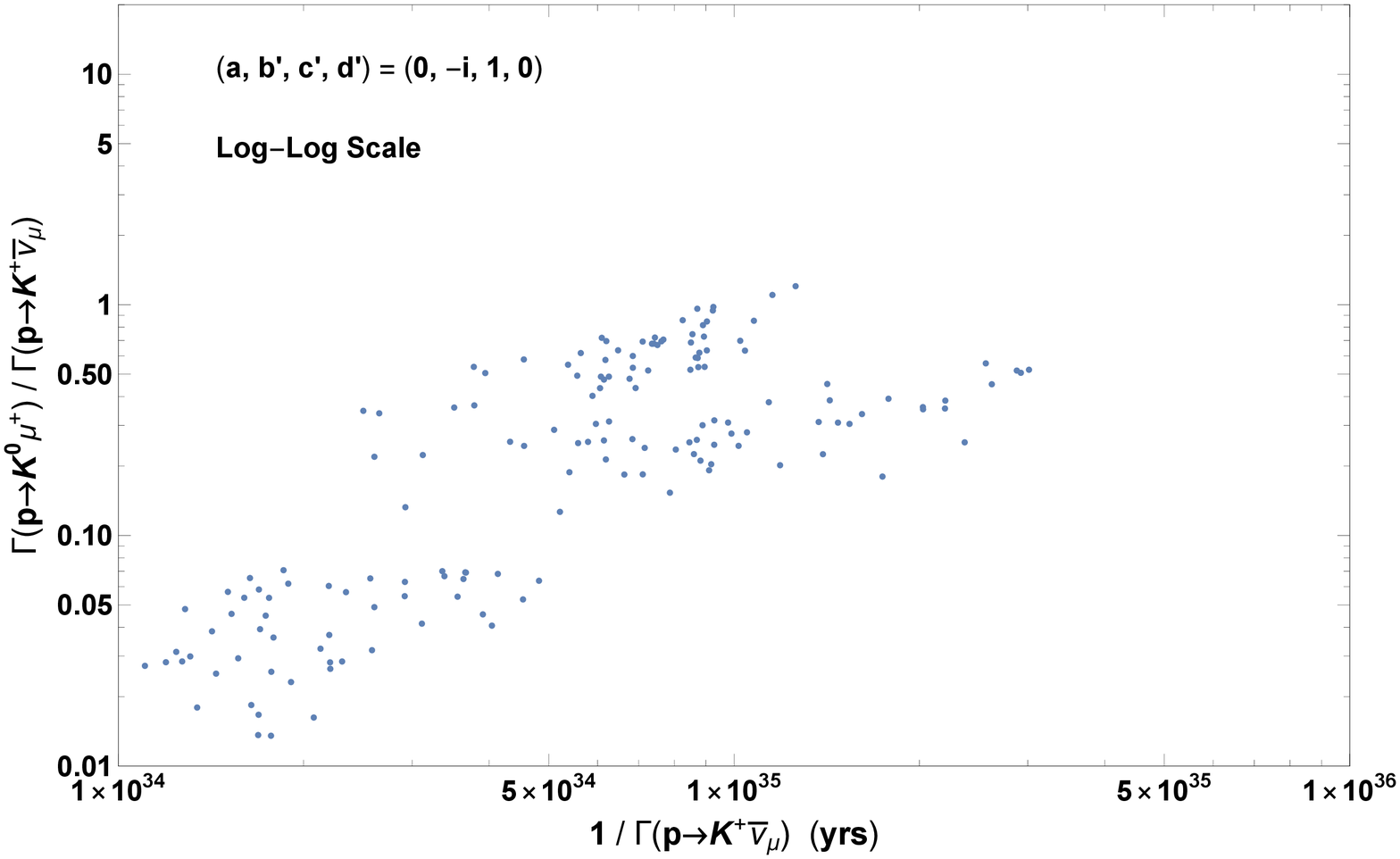}
\caption{
Same as Fig.~\ref{plot-indiv} except that
 we take $(a,b',c',d')=(0,1,1,0)$, $(0,i,1,0)$, $(0,-1,1,0)$, $(0,-i,1,0)$ in Eq.~(\ref{yyl1-alt2})-(\ref{yyl3-alt2}).
}
\label{plot-cb}
\end{center}
\end{figure}
We observe that {\it when $d'=0$, $c'=O(1)$ and $b'=O(1)$, 
 the prediction on the partial width ratio varies considerably with the relative phase of $b'$ and $c'$
 and with different fitting results.}
Still, we can assert that the ratio is above 0.01.
The absence of strong suppression factor $0.3\cdot0.002$ is consistent with our estimate Eq.~(\ref{ratioso10}).

\item
When $d'=b'=0$, both $\Gamma(p\to K^+\bar{\nu}_\mu)$ and $\Gamma(p\to K^0\mu^+)$ are
 dominated by the contribution from the term with $c'$.
We thus conclude that {\it when $d'=b'=0$, irrespectively of the value of $a$, the prediction on the partial width ratio is given 
 by the lower-left panel of Fig.~\ref{plot-indiv},
 where it varies in the ranges 0.03-0.2 and 0.4-0.8.}

\item
When $d'=c'=0$, the partial width $\Gamma(p\to K^+\bar{\nu}_\mu)$ is dominated by the contribution from
 the term with $b'$.
On the other hand, since the partial width ratio $\Gamma(p\to K^0\mu^+)/\Gamma(p\to K^+\bar{\nu}_\mu)$
  is much larger with $(a,b',c',d')=(1,0,0,0)$ than with $(a,b',c',d')=(0,1,0,0)$,
 $\Gamma(p\to K^0\mu^+)$ might receive larger contribution from the term with $a$ than from the term with $b'$.
However, we have inspected cases with $(a,b',c',d')=(1,1,0,0)$, $(i,1,0,0)$, $(-1,1,0,0)$, $(-i,1,0,0)$
 and found that the distribution in these cases is almost identical to that with $(a,b',c',d')=(0,1,0,0)$.
We thus conclude that {\it when $d'=c'=0$, irrespectively of the value of $a$,
 the prediction on the partial width ratio is given by the upper-right panel of Fig.~\ref{plot-indiv},
 where it is mostly suppressed below 0.0005.}
This result agrees with our estimate Eq.~(\ref{ratioso10-finetuned}).

\item
Only in the very special case with $d'=c'=b'=0$ do we obtain the distribution of the upper-left panel of Fig.~\ref{plot-indiv},
 where the ratio is above 0.05.

\end{itemize}
To summarize, if $d'=O(1)$, 
 the partial width ratio $\Gamma(p\to K^0\mu^+)/\Gamma(p\to K^+\bar{\nu}_\mu)$ is mostly in the range 0.05-0.6.
If $d'=0$, $c'=O(1)$ and $b'=O(1)$, the partial width ratio varies in a wide range, still it is above 0.01.
If $d'=b'=0$ and $c'=O(1)$, it is in the ranges 0.03-0.2 and 0.4-0.8.
If $d'=c'=b'=0$, it is above 0.05.
Only when $d'=c'=0$ and $b'=O(1)$ is the partial width ratio mostly highly suppressed below 0.0005.

Because there is no particular reason to believe $d'=0$,
 our most important result is the lower-right panel of Fig.~\ref{plot-indiv}, which covers the case with $d'=O(1)$.
Accordingly, our main prediction is
\bea
0.6 \ \gtrsim \ \frac{\Gamma(p\to K^0\mu^+)}{\Gamma(p\to K^+\bar{\nu}_\mu)} \ \gtrsim \ 0.05.
\eea
Considering the current 90\% CL bound $1/\Gamma(p\to K^+\nu) > 5.9\times10^{33}$~years~\cite{Abe:2014mwa},
 we can at best observe the $p\to K^0\mu^+$ decay at a rate $1/\Gamma(p\to K^0\mu^+) =1\times 10^{34}$~years.
\\

\section{Summary}

The ratio of the partial widths of some dimension-5 proton decay modes can be predicted without knowledge of SUSY particle masses,
 and thus serves as a probe for various SUSY GUT models even when SUSY particles are not discovered.
We have focused on the partial width ratio $\Gamma(p\to K^0\mu^+)/\Gamma(p\to K^+\bar{\nu}_\mu)$
 in the minimal renormalizable SUSY $SO(10)$ GUT.
In the model, the Wilson coefficients of dimension-5 operators responsible for the $p\to K^0\mu^+$ and the $p\to K^+\bar{\nu}_\mu$
 decays are on the same order, and
 $\Gamma(p\to K^0\mu^+)/\Gamma(p\to K^+\bar{\nu}_\mu)$ is largely
 determined by the ratio of baryon chiral Lagrangian parameters and is estimated to be $O(0.1)$.
This is in striking contrast to the minimal $SU(5)$ GUT, where this partial width ratio is further suppressed by factor
 $y_u^2/(\lambda^2 y_c)^2\simeq0.002$.
To confirm that $\Gamma(p\to K^0\mu^+)/\Gamma(p\to K^+\bar{\nu}_\mu)=O(0.1)$ in the minimal renormalizable SUSY $SO(10)$ GUT,
 we have numerically determined $Y_{10},Y_{126}$ through a fitting of the quark and charged lepton Yukawa couplings and neutrino mass matrix,
 and calculated the partial width ratio based on the fitting results.
Our most important finding is that the partial width ratio generally varies in the range
 $0.6\gtrsim\Gamma(p\to K^0\mu^+)/\Gamma(p\to K^+\bar{\nu}_\mu)\gtrsim0.05$
 in the most generic case where $d'=O(1)$ in Eqs.~(\ref{yyl1-alt2})-(\ref{yyl3-alt2}).
\\

\section*{Acknowledgement}
This work is partially supported by Scientific Grants by the Ministry of Education, Culture, Sports, Science and Technology of Japan,
Nos.~17K05415, 18H04590 and 19H051061 (NH), and No.~19K147101 (TY).
\\


\end{document}